\documentclass[aps,prl,reprint,superscriptaddress,nofootinbib,longbibliography,floatfix]{revtex4-2}

\usepackage{amsmath,amssymb,bm}
\usepackage{graphicx}
\usepackage{orcidlink}
\usepackage{hyperref}
\hypersetup{hidelinks}

\newcommand{\bk}{\bm k}
\newcommand{\bq}{\bm q}
\newcommand{\be}{\bm e}
\newcommand{\angstrom}{\text{\normalfont\AA}}

\begin{document}

\title{Rank-Selective Optical Tomography of Higher-Wave Altermagnetism}

\author{Meysam Bagheri Tagani\orcidlink{0000-0001-7065-3195}}
\email{mtagani@magtop.ifpan.edu.pl}
\affiliation{Department of Physics, University of Guilan, P. O. Box 41335-1914, Rasht, Iran}
\affiliation{International Research Centre MagTop, Institute of Physics, Polish Academy of Sciences, Aleja Lotnik\'ow 32/46, PL-02668 Warsaw, Poland}

\author{Carmine Autieri\orcidlink{0000-0002-5008-8165}}
\email{autieri@magtop.ifpan.edu.pl}
\affiliation{International Research Centre MagTop, Institute of Physics, Polish Academy of Sciences, Aleja Lotnik\'ow 32/46, PL-02668 Warsaw, Poland}

\author{Sahar Izadi Vishkayi\orcidlink{0000-0002-0314-1007}}
\email{izadi.123@gmail.com}
\affiliation{International Research Centre MagTop, Institute of Physics, Polish Academy of Sciences, Aleja Lotnik\'ow 32/46, PL-02668 Warsaw, Poland}

\date{\today}

\begin{abstract}
Identifying the spatial rank of higher-wave altermagnetic order optically is challenging because local electric-dipole response does not uniquely resolve distinct continuum harmonics. We show that finite photon momentum turns one-photon spin-resolved absorption into a rank-selective tomography. For the planar $|m|=\ell$ sector of an even-parity $\ell$-wave component, a joint Fourier projection in polarization and momentum angle isolates
$\mathcal T_\ell\propto\eta_\ell q^{\ell-2}$, yielding the hierarchy
$d:q^0$, $g:q^2$, and $i:q^4$; phase changes track rotations of the selected magnetic harmonic. A Ward-consistent finite-$q$ microscopic calculation reproduces these powers without imposing them. Although discrete crystal symmetry can generate lower-order local aliases, they are orthogonal to the selected momentum harmonic and cannot contaminate it below $q^{\ell-2}$. In MnTe, whose nonrelativistic parent order is three-dimensional $g$ wave while spin--orbit coupling lowers the exact relativistic spin-momentum-locking symmetry, first-principles calculations show that more than $99.9\%$ of the Fourier power of the N\'eel-projected $A$-region spin-energy contrast remains in the parent $g$-wave-derived $m=3$ harmonic. Structured near fields place the required momentum window within experimental reach.
\end{abstract}

\maketitle

Altermagnets combine compensated collinear order with non-relativistic, momentum-dependent spin splitting, whose even-parity form factor is classified as $d$-, $g$-, or $i$-wave~\cite{Smejkal2022PRX,Jungwirth2026Nature,McClarty2024PRL,Bhowal2024PRX}. Spin-resolved spectroscopy has established that these higher-wave textures are experimentally accessible~\cite{Lee2024PRL,Ding2024PRL,Jiang2025NatPhys}.  Their defining information, however, is an angular rank and nodal pattern in momentum space, whereas conventional optical response is usually evaluated in the local electric-dipole limit.
Recent work accesses altermagnetic consequences through orbital and quantum geometry, elliptic dichroism, nonlinear multipoles, multiphoton absorption, magnon Raman response, or real-space N\'eel textures~\cite{Vila2025PRB,Li2026PRL,Ezawa2026PRB,Sunko2026npj,Usachev2026PRL,Ghorashi2026arXiv,Yuan2026PRL,Maiani2026PRL}.  Spin-space-group analyses have further shown that non-relativistic symmetries can constrain optical tensors beyond magnetic-group predictions~\cite{Sivianes2026arXiv}.  None of these results provides a one-photon linear observable whose photon-momentum harmonic directly measures the missing spatial rank of a uniform higher-wave order. Here, that rank is supplied by the magnitude and direction of photon momentum.  We derive a rank-matching rule, construct a joint angular projection immune to lower-order crystalline aliases, verify it microscopically, and identify a feasible near-field protocol.  First-principles results for MnTe further show that the parent nonrelativistic \(g\)-wave-derived component remains strongly encoded in the Néel-projected electronic texture near \(A\) after SOC.

\textit{Rank-selective spatial dispersion.---}
A pure even-parity component has compensated spin splitting $\Delta_\ell(\bk)\propto\Phi_\ell(\bk)$ with $\ell=2,4,6$ for $d$-, $g$-, and $i$-wave order.  We consider finite-$\bq$ absorption resolved along a spin axis $\hat{\bm n}$ and retain the part odd under reversal of all local moments,
\begin{equation}
\Delta W_{\hat{\bm n}}
=\left[W_{+\hat{\bm n}}-W_{-\hat{\bm n}}\right]_{\rm odd}
=n_a e_i^*\Xi^{\rm odd}_{a;ij}(\bq,\omega)e_j .
\label{eq:spincontrast}
\end{equation}
The finite-$\bq$ transition rate, spin projection, and gauge-covariant formulation are given in the Supplemental Material (SM)~\cite{SM,Pozo2023SciPost}.

\begin{figure*}[t]
\centering
\includegraphics[width=\linewidth]{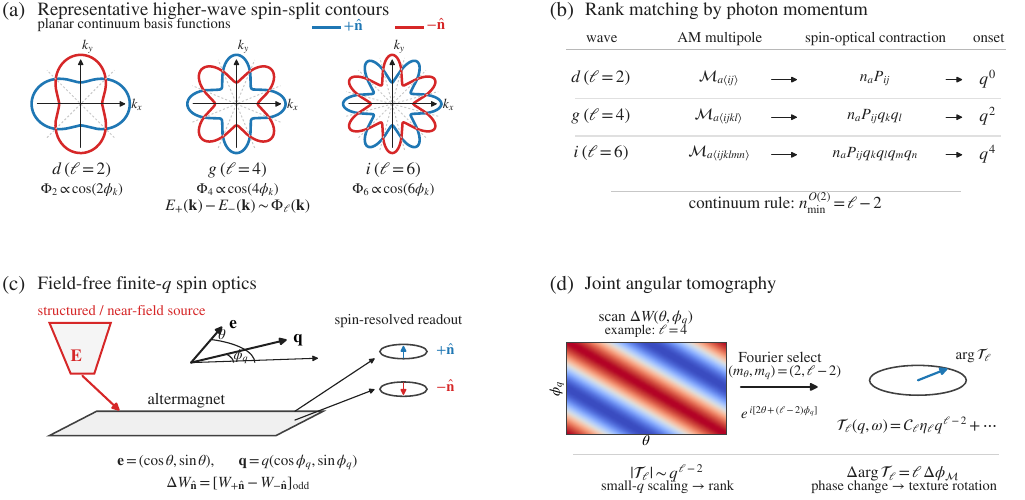}
\caption{\label{fig:concept}
Rank-selective optical tomography.  Panel (a) shows representative two-dimensional continuum basis functions used to illustrate the rank-matching principle; material-specific three-dimensional \(g\)-wave harmonics may have a different in-plane winding.  (b) Linear polarization supplies spatial rank two, while photon momentum supplies the remaining rank, giving $d:q^0$, $g:q^2$, and $i:q^4$.  (c) Field-free finite-$\bq$ geometry.  (d) A joint angular projection isolates $\mathcal T_\ell\propto\eta_\ell q^{\ell-2}$; the small-\(q\) scaling of its magnitude identifies the spatial-dispersion order, while phase changes track the relative orientation of the magnetic harmonic.}
\end{figure*}

A pure $\ell$-wave component may be represented by a symmetric-traceless tensor $\mathcal M_{a\langle i_1\cdots i_\ell\rangle}$; the rank-two linear-polarization tensor $P_{ij}=\mathrm{Re}(e_i^*e_j)-\delta_{ij}|\be|^2/2$ then gives the leading rotationally resolved invariant
\begin{align}
\Delta W_\ell^{\rm LD}
&=C_\ell(\omega)n_a\mathcal M_{a\langle i_1\cdots i_\ell\rangle}
P_{i_1i_2}q_{i_3}\cdots q_{i_\ell}+\mathcal O(q^\ell),
\nonumber\\[-2pt]
n_{\min}^{O(2)}&=\ell-2,
\qquad d:q^0,\quad g:q^2,\quad i:q^4 .
\label{eq:rankmatching}
\end{align}
The second line follows because an order-$q^n$ polynomial contains no momentum-angle harmonic with $|m_q|>n$, whereas $P_{ij}$ already supplies $|m|=2$. Equation~\eqref{eq:rankmatching} is a tensor statement, whereas the angular representation below specializes to a planar \(|m|=\ell\) component; three-dimensional crystalline components obey the same rank counting but generally have a material- and geometry-dependent angular projector (SM).

Writing $\be=(\cos\theta,\sin\theta)$ and $\bq=q(\cos\phi_q,\sin\phi_q)$, the rank-resolved component is isolated by
\begin{align}
\mathcal T_\ell(q,\omega)
&=\int\!\frac{d\theta\,d\phi_q}{(2\pi)^2}
 e^{i[2\theta+(\ell-2)\phi_q]}
\Delta W_{\hat{\bm n}}(q;\theta,\phi_q)
\nonumber\\
&=\mathcal C_\ell(\omega)\eta_\ell q^{\ell-2}+\mathcal O(q^\ell),
\qquad
\eta_\ell=|\eta_\ell|e^{i\ell\phi_{\mathcal M}} .
\label{eq:tomography}
\end{align}
Thus, the small-$q$ power identifies the selected spatial rank, while phase changes track rotations of the magnetic texture.  The degree bound $|m_q|\le n$ protects this Fourier sector against every contribution with $n<\ell-2$, independent of crystal symmetry.

A discrete crystal can nevertheless permit lower-order terms in the total response because continuum angular momenta subduce to the same point-group representation.  We call such terms crystalline aliases.  They can produce, for example, a local $g$-wave optical signal~\cite{Haag2026PRM}, but a $q^0$ alias has $m_q=0$ and is orthogonal to the $m_q=2$ projection defining $\mathcal T_4$.  The exact representation criterion, mirror and inversion constraints, and the general mod-$N$ aliasing rule are derived in the SM.  Unlike field- or strain-assisted birefringence and gradient-assisted second-harmonic generation~\cite{Sunko2026npj,Usachev2026PRL}, the analyzed spin channel supplies the axial index and photon momentum supplies the missing spatial rank at fixed optical-field order.

\textit{Microscopic realization and crystalline protection.---}
We test the selection rule with the rank-pure continuum model
\begin{equation}
\begin{aligned}
H_\ell(\bk)&=\hbar v(k_x\rho_z\tau_x+k_y\tau_y)
+\left[m+\Delta_\ell\Phi_\ell(\bk)s_z\right]\tau_z,\\
\Phi_\ell(\bk)&=k_0^{-\ell}\mathrm{Re}
\!\left[e^{-i\ell\phi_{\mathcal M}}(k_x+ik_y)^\ell\right].
\end{aligned}
\label{eq:model}
\end{equation}
Here $\tau_i$, $s_i$, and $\rho_i$ act in orbital, spin, and time-reversed
flavor spaces.  $k_0$ is a reference momentum scale that renders
$\Phi_\ell$ dimensionless, with $E_0=\hbar v k_0$ used as the
corresponding energy unit.  For even $\ell$, inversion is preserved, time reversal is
broken, and
$E_{\nu,+}(\bk)=E_{\nu,-}(R_{\pi/\ell}\bk)$.
The two spin sectors, therefore, have symmetry-related integrated spectra
while remaining split at generic momentum, providing a compensated
rank-pure higher-wave benchmark without SOC.  At finite momentum, we evaluate the full shifted-state interband response with the straight-line
current vertex given in the SM.  The construction obeys the longitudinal
Ward identity exactly and is important here because the $g$- and $i$-wave
form factors contain high powers of momentum; using only the local
$\partial_{\bk}H$ vertex would not provide a controlled finite-$q$
benchmark.

\begin{figure*}[t]
\centering
\includegraphics[width=\linewidth]{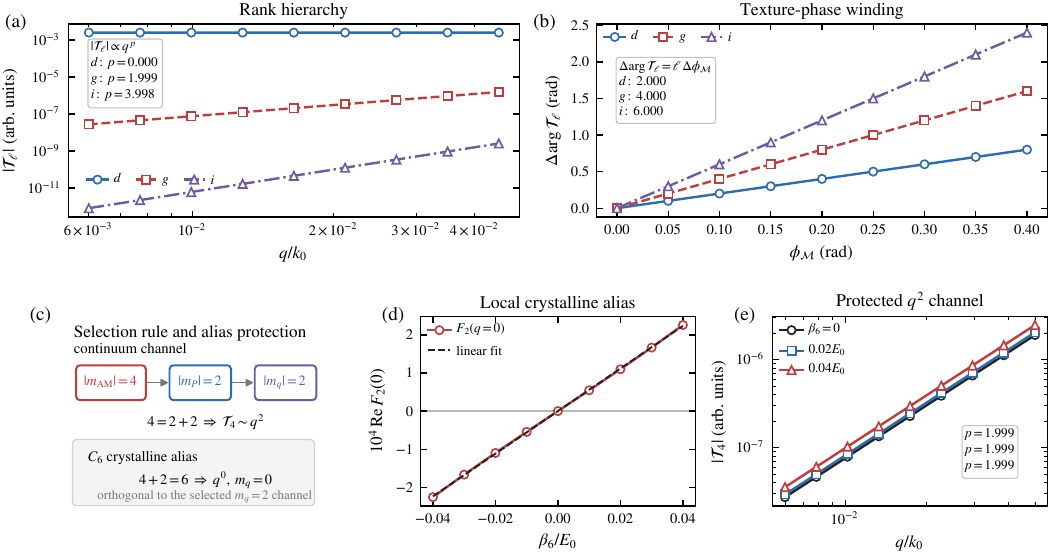}
\caption{\label{fig:validation}
Microscopic rank hierarchy and crystalline protection.  (a) The projected
response scales as $q^0$, $q^2$, and $q^4$ for $d$-, $g$-, and $i$-wave
order.  (b) Rotating the texture produces phase windings $2$, $4$, and $6$.
(c) Rank bookkeeping for the $g$-wave channel and its $C_6$ local alias.
(d) The local \(m_\theta=2\) linear-polarization harmonic \(F_2(q=0)\) is linear in the hexagonal anisotropy \(\beta_6\).  (e) Despite this alias, the projected
$\mathcal T_4$ retains its $q^2$ onset.  Parameters and convergence tests
are given in the SM.}
\end{figure*}

\begin{figure*}[t]
\centering
\includegraphics[width=\linewidth]{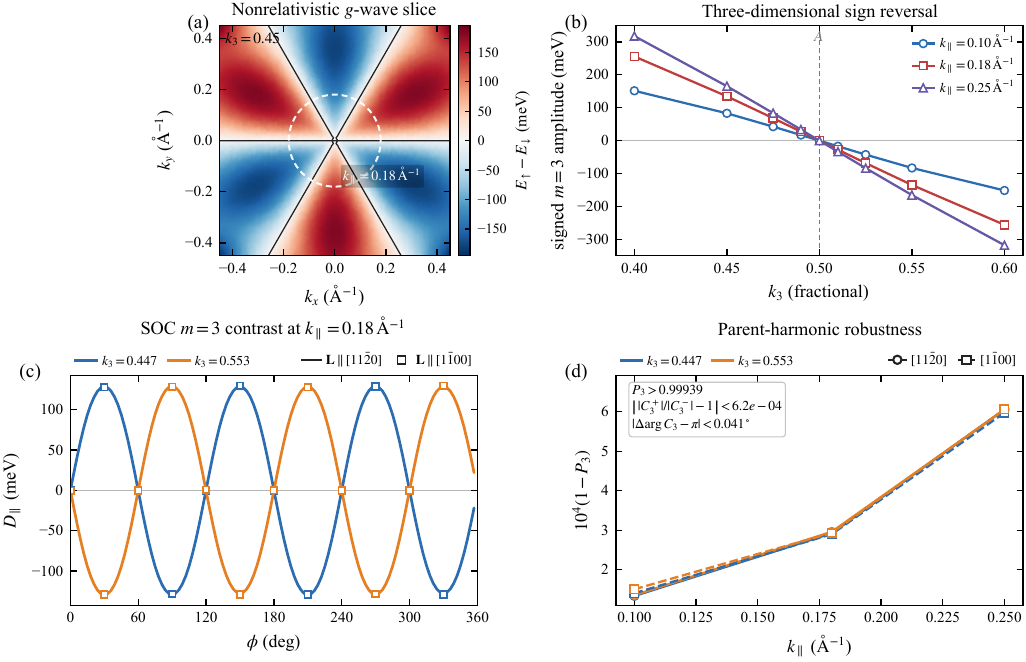}
\caption{\label{fig:mnte}
First-principles higher-wave texture in MnTe.
(a) Nonrelativistic spin splitting on the $k_3=0.45$ plane.
The white dashed circle marks the
$k_\parallel=0.18\,\mathrm{\AA}^{-1}$ contour highlighted in panel (c).
(b) Signed $m=3$ Fourier amplitude across the $A$ plane for three
in-plane radii, showing its zero and sign reversal at $A$.
(c) N\'eel-projected spin-energy contrast $D_\parallel$ including SOC
at $k_\parallel=0.18\,\mathrm{\AA}^{-1}$ for the two planes flanking
$A$ and two in-plane N\'eel orientations.
(d) Residual Fourier power $1-P_3$ outside the parent $m=3$ harmonic
of $D_\parallel$, demonstrating the robustness of the parent
higher-wave component under SOC.
}
\end{figure*}

No power law is imposed numerically.  Figure~\ref{fig:validation}(a)
gives $p_2\simeq0$, $p_4\simeq2$, and $p_6\simeq4$ for
$|\mathcal T_\ell|\propto q^{p_\ell}$, so the symmetry hierarchy emerges
directly from the interband dynamics.  Independently,
Fig.~\ref{fig:validation}(b) gives
$\Delta\arg\mathcal T_\ell=\ell\,\delta\phi_{\mathcal M}$.
The magnitude and phase, therefore, encode complementary information:
the small-$q$ exponent identifies the selected spatial rank, whereas phase
changes measure the rotation of the higher-wave texture without requiring
knowledge of the nonuniversal phase of $\mathcal C_\ell(\omega)$.
The selection rule fixes the leading allowed power, not its spectral
weight; additional symmetries or accidental zeros can suppress
$\mathcal C_\ell(\omega)$ without changing the rank criterion.

We next test the more demanding case in which the crystal itself produces
a lower-order optical signal.  Adding the time-reversal-even hexagonal mass
harmonic $\beta_6\Psi_6\tau_z$ to the $g$-wave model generates the
crystalline alias allowed by $C_6$.  Such a $k^6\cos6\phi_k$ structural
term also appears in a symmetry-derived MnTe Hamiltonian
~\cite{Takahashi2025npj}. Note that the \(C_6\) model is a generic planar rank-four benchmark rather than a material-specific representation of the MnTe magnetic basis; its connection to MnTe here concerns the crystal-allowed sixfold structural harmonic.
Because $m_{\rm AM}=4$ and $m_P=2$ can close through the structural
$m_C=6$ harmonic, a domain-odd $q^0$ polarization signal develops and is
linear in $\beta_6$ at weak anisotropy
[Fig.~\ref{fig:validation}(c,d)].  Crucially, we do not assume this local
alias to be small or absent: the $m_q=2$ Fourier projection removes it,
and the calculated $\mathcal T_4$ retains its $q^2$ onset for every
anisotropy studied [Fig.~\ref{fig:validation}(e)].  Thus, the protection is
stronger than a statement about an ideal rotationally invariant model; it
survives in the presence of an explicitly generated crystal-allowed local
response.  The SM gives the corresponding unprojected Fourier spectra and
an independent $i$-wave/$C_4$ example, where a $q^0$ alias likewise
coexists with a protected $\mathcal T_6\propto q^4$ channel.

\textit{First-principles realization in MnTe.---}
As a material testbed, we use hexagonal MnTe, a representative bulk $g$-wave
altermagnet~\cite{Lee2024PRL}. We first suppress SOC to expose the
non-relativistic exchange texture. The calculated state is compensated,
with opposite Mn moments of approximately $4.4\,\mu_{\rm B}$ and vanishing
net magnetization.  On the $k_3=0.45$ plane, the top valence band displays
the six-sector pattern of Fig.~\ref{fig:mnte}(a), with essentially all
angular weight concentrated in the $m=3$ Fourier harmonic.  The apparently
threefold in-plane object is nevertheless a three-dimensional $g$-wave
texture: its coefficient vanishes at the $A$ plane and obeys
$C_3(1/2+\delta)\simeq-C_3(1/2-\delta)$
[Fig.~\ref{fig:mnte}(b)].  Near \(A\), this behavior is consistent with a leading local form \(\delta k_z\,\mathrm{Re}[e^{-i3\phi_0}(k_x+ik_y)^3]\), with \(\delta k_z\propto k_3-\tfrac12\), whose total momentum degree is four.  The sign reversal, therefore, supplies
the missing out-of-plane factor that converts the $m=3$ slice harmonic into
the three-dimensional $\ell=4$ structure of the bulk $g$-wave state.

SOC removes exact spin conservation, so the robustness test must not rely on
assigning the two relativistic branches fixed $\uparrow/\downarrow$ labels.
For the upper and lower members $v_{\rm u,l}$ of the top valence pair we
therefore define the N\'eel-projected spin-energy contrast
\begin{equation}
\begin{aligned}
D_\parallel(\bk)&=\tfrac12\Delta E_v(\bk)\Delta s_\parallel(\bk),\\
\Delta E_v&=E_{v_{\rm u}}-E_{v_{\rm l}},\qquad
\Delta s_\parallel=s_{v_{\rm u},\parallel}-s_{v_{\rm l},\parallel}.
\end{aligned}
\label{eq:Dparallel}
\end{equation}
Here $s_{v,\parallel}$ is the normalized PAW-projected spin polarization
along the N\'eel vector.  In the spin-conserving limit, this quantity reduces
to the signed exchange splitting, while after SOC, it continuously weights the band separation by the surviving spin contrast. Note that the in-plane Néel vector lowers the exact relativistic spin-momentum-locking symmetry of MnTe\cite{autieri2026relativistic}. Therefore, \(P_3\) should not be interpreted as a relativistic symmetry classification; it quantifies instead the local Fourier dominance of the parent nonrelativistic \(g\)-wave-derived component in the Néel-projected \(A\)-region spin-energy contrast.  We evaluate two
inequivalent orientations,
$\bm L\parallel[11\bar{2}0]$ and
$\bm L\parallel[1\bar{1}00]$.

The resulting contrast makes the relativistic robustness particularly
clear.  SOC strongly reshapes the valence topology near $\Gamma$ (SM), yet the $A$-region higher-wave pattern is almost unchanged
[Fig.~\ref{fig:mnte}(c)].  Across
$k_\parallel=0.10$, $0.18$, and $0.25\,\angstrom^{-1}$ on the two planes
$k_3=0.447,0.553$, the $m=3$ Fourier power satisfies
$P_3>0.99939$, so less than $6.1\times10^{-4}$ of the angular power leaks
into all other harmonics.  Opposite-plane amplitudes agree within
$6.2\times10^{-4}$, while their phase differs from $\pi$ by less than
$0.041^\circ$ [Fig.~\ref{fig:mnte}(d)].  Rotating the in-plane N\'eel
vector changes the $m=3$ amplitude only at the $10^{-2}\%$ level.
The coexistence of strong SOC-induced reconstruction elsewhere in the Brillouin zone with this exceptionally stable $A$-region harmonic shows
that the persistence is nontrivial rather than a consequence of negligible
relativistic coupling.  These calculations do not evaluate the finite-\(\mathbf q\) optical coefficient itself; rather, they establish that MnTe provides a realistic three-dimensional rank-four parent texture whose \(g\)-wave-derived \(m=3\) component remains strongly encoded in the Néel-projected \(A\)-region electronic structure after SOC.

\textit{Feasibility and outlook.---}
A structured near-field or polaritonic mode supplies
$q\simeq2\pi/\Lambda$.  Confinement periods
$\Lambda=10$--$100$~nm therefore correspond to
$q\simeq6\times10^{-3}$--$6\times10^{-2}\,\angstrom^{-1}$, reaching
$q/k_0\sim10^{-2}$--$10^{-1}$ for representative electronic momentum
scales~\cite{Guo2025LSA,Hao2026RSI}.  This enhancement is essential for
higher-rank order: after factoring out the nonuniversal spectral coefficient,
the purely kinematic suppression is $(q/k_0)^{\ell-2}$, so at
$q/k_0=0.1$ the $d$-, $g$-, and $i$-wave factors are respectively
$1$, $10^{-2}$, and $10^{-4}$.  These numbers are not absolute absorption
fractions---resonant matrix elements, near-field enhancement, temperature,
and broadening remain material-dependent --- but they quantify the momentum
penalty that an experiment must overcome.

The tomography requires angular control rather than a continuum of
measurements.  After domain-odd and counterpropagating-$q$ symmetrization, an ideal band-limited reconstruction uses three linear-polarization axes
and $1$, $3$, or $5$ distinct momentum axes for $d$-, $g$-, or $i$-wave
order, respectively; oversampling can reject higher crystalline harmonics
and improve statistics.  Finite angular acceptance changes the amplitude
but not the selected Fourier index: for a Gaussian rms momentum-angle
spread of $10^\circ$, the ideal $g$- and $i$-wave harmonics retain about
$94\%$ and $78\%$ of their perfectly collimated amplitudes (SM).
The analyzed spin channel may be read directly by spin-resolved
photoemission~\cite{deLaFiguera2017} or transduced into a spin-selective
photocurrent, while the optical momentum launcher and spin detector remain
independent experimental components.

The rank can therefore be cross-validated in three independent ways: the
power of $q$ fixes the spatial-dispersion order, the momentum-angle Fourier
index selects the corresponding rank sector, and the phase winding under
texture rotation measures $\ell$ independently.  This combination is what
makes the protocol robust when local crystal-allowed optical signals
coexist.  Our result is therefore a one-photon, field-free spectroscopy
theorem rather than a material-specific dichroic fingerprint.  It is
complementary to multiphoton hierarchies~\cite{Ghorashi2026arXiv} and to
dichroism generated by spatially textured N\'eel order
~\cite{Maiani2026PRL}: here, the optical field order and magnetic texture
remain fixed, while the continuously controlled photon momentum supplies
and resolves the missing spatial rank.  The same construction can be
applied more broadly whenever a controllable probe momentum supplies the tensor structure that is absent from the local response.\\

\begin{acknowledgments}
M.B.T. acknowledges support from the Polish National Agency for Academic Exchange (NAWA) under the ULAM program, project BPN/ULM/2025/1/00156/U/00001, and from the Iran National Science Foundation under project 4043973.  This work was supported by the Foundation for Polish Science project ``MagTop,'' FENG.02.01-IP.05-0028/23, co-financed by the European Union under the European Funds for a Smart Economy Program 2021--2027.  We further acknowledge access to the computing facilities of the Interdisciplinary Center of Modeling at the University of Warsaw, Grant g91-1418, g91-1419, g96-1808, g96-1809 and g103-2540 for the availability of high-performance computing resources and support. We acknowledge access to the computing facilities of the Poznan Supercomputing and Networking Center, Grants No. pl0267-01, pl0365-01, pl0807, and pl0471-01.
\end{acknowledgments}

\section*{Data availability}
The numerical data and analysis scripts supporting this work are available from the corresponding authors upon reasonable request.

\nocite{Sato2026npj,Kresse1996VASP,Kresse1999VASP,Blochl1994PAW,PBE1996,Liechtenstein1995,Monkhorst1976,deLaFiguera2017,Gunyaga2023PRB,Gunyaga2026PRB}
\bibliography{references_prl_final}

@article{Smejkal2022PRX,
  author  = {Libor {\v S}mejkal and Jairo Sinova and Tomas Jungwirth},
  title   = {Beyond Conventional Ferromagnetism and Antiferromagnetism: A Phase with Nonrelativistic Spin and Crystal Rotation Symmetry},
  journal = {Phys. Rev. X},
  volume  = {12},
  pages   = {031042},
  year    = {2022},
  doi     = {10.1103/PhysRevX.12.031042}
}

@article{Jungwirth2026Nature,
  author  = {Tomas Jungwirth and Jairo Sinova and Rafael M. Fernandes and Qihang Liu and Hikaru Watanabe and Shuichi Murakami and Satoru Nakatsuji and Libor {\v S}mejkal},
  title   = {Symmetry, microscopy and spectroscopy signatures of altermagnetism},
  journal = {Nature},
  volume  = {649},
  pages   = {837--847},
  year    = {2026},
  doi     = {10.1038/s41586-025-09883-2}
}

@article{McClarty2024PRL,
  author  = {Paul A. McClarty and Jeffrey G. Rau},
  title   = {Landau Theory of Altermagnetism},
  journal = {Phys. Rev. Lett.},
  volume  = {132},
  pages   = {176702},
  year    = {2024},
  doi     = {10.1103/PhysRevLett.132.176702}
}

@article{Bhowal2024PRX,
  author  = {Sayantika Bhowal and Nicola A. Spaldin},
  title   = {Ferroically Ordered Magnetic Octupoles in {$d$}-Wave Altermagnets},
  journal = {Phys. Rev. X},
  volume  = {14},
  pages   = {011019},
  year    = {2024},
  doi     = {10.1103/PhysRevX.14.011019}
}

@article{Lee2024PRL,
  author  = {Suyoung Lee and Sangjae Lee and Saegyeol Jung and Jiwon Jung and Donghan Kim and Yeonjae Lee and Byeongjun Seok and Jaeyoung Kim and Byeong Gyu Park and Libor {\v S}mejkal and Chang-Jong Kang and Changyoung Kim},
  title   = {Broken Kramers Degeneracy in Altermagnetic {MnTe}},
  journal = {Phys. Rev. Lett.},
  volume  = {132},
  pages   = {036702},
  year    = {2024},
  doi     = {10.1103/PhysRevLett.132.036702}
}

@article{Ding2024PRL,
  author  = {Jianyang Ding and Zhicheng Jiang and Xiuhua Chen and Zicheng Tao and Zhengtai Liu and others},
  title   = {Large Band Splitting in {$g$}-Wave Altermagnet {CrSb}},
  journal = {Phys. Rev. Lett.},
  volume  = {133},
  pages   = {206401},
  year    = {2024},
  doi     = {10.1103/PhysRevLett.133.206401}
}

@article{Jiang2025NatPhys,
  author  = {Bei Jiang and Mingzhe Hu and Jianli Bai and Ziyin Song and Chao Mu and others},
  title   = {A metallic room-temperature {$d$}-wave altermagnet},
  journal = {Nat. Phys.},
  volume  = {21},
  pages   = {754--759},
  year    = {2025},
  doi     = {10.1038/s41567-025-02822-y}
}

@article{Vila2025PRB,
  author  = {Marc Vila and Veronika Sunko and Joel E. Moore},
  title   = {Orbital-spin locking and its optical signatures in altermagnets},
  journal = {Phys. Rev. B},
  volume  = {112},
  pages   = {L020401},
  year    = {2025},
  doi     = {10.1103/bzzy-ngcs}
}

@article{Li2026PRL,
  author  = {Yongpan Li and Cheng-Cheng Liu},
  title   = {Quantum-Metric-Based Optical Selection Rules},
  journal = {Phys. Rev. Lett.},
  volume  = {136},
  pages   = {046901},
  year    = {2026},
  doi     = {10.1103/bdhy-hnd2}
}

@article{Ezawa2025PRB,
  author  = {Motohiko Ezawa},
  title   = {Third-order and fifth-order nonlinear spin-current generation in {$g$}-wave and {$i$}-wave altermagnets and perfectly nonreciprocal spin current in {$f$}-wave magnets},
  journal = {Phys. Rev. B},
  volume  = {111},
  pages   = {125420},
  year    = {2025},
  doi     = {10.1103/PhysRevB.111.125420}
}

@article{Haag2026PRM,
  author  = {Luca Felipe Haag and Marius Weber and Kai Leckron and Libor {\v S}mejkal and Jairo Sinova and Hans Christian Schneider},
  title   = {Signatures of bulk {$g$}-wave altermagnetism in optically excited {$\alpha$-MnTe}},
  journal = {Phys. Rev. Mater.},
  volume  = {10},
  pages   = {074413},
  year    = {2026},
  doi     = {10.1103/ypfs-fy39}
}

@article{Sunko2026npj,
  author  = {Veronika Sunko and Joseph Orenstein},
  title   = {Linear magneto-birefringence as a probe of altermagnetism},
  journal = {npj Quantum Mater.},
  year    = {2026},
  doi     = {10.1038/s41535-026-00901-8}
}

@article{Usachev2026PRL,
  author  = {P. A. Usachev and R. V. Pisarev and V. V. Pavlov},
  title   = {Nonlinear optical probing of ferroic octupolar order parameter in collinear altermagnet},
  journal = {Phys. Rev. Lett.},
  year    = {2026},
  note    = {Accepted 7 August 2026},
  doi     = {10.1103/h1ht-vsrn}
}

@article{Pozo2023SciPost,
  author  = {{\'O}scar Pozo Oca{\~n}a and Ivo Souza},
  title   = {Multipole theory of optical spatial dispersion in crystals},
  journal = {SciPost Phys.},
  volume  = {14},
  pages   = {118},
  year    = {2023},
  doi     = {10.21468/SciPostPhys.14.5.118}
}

@article{Gunyaga2023PRB,
  author  = {A. A. Gunyaga and M. V. Durnev and S. A. Tarasenko},
  title   = {Photocurrents induced by structured light},
  journal = {Phys. Rev. B},
  volume  = {108},
  pages   = {115402},
  year    = {2023},
  doi     = {10.1103/PhysRevB.108.115402}
}

@article{Gunyaga2026PRB,
  author  = {A. A. Gunyaga and M. V. Durnev and S. A. Tarasenko},
  title   = {Electric and spin-valley currents induced by structured light in two-dimensional Dirac materials},
  journal = {Phys. Rev. B},
  volume  = {113},
  pages   = {075423},
  year    = {2026},
  doi     = {10.1103/sbhz-2nvs}
}

@article{Hao2026RSI,
  author  = {Duxing Hao and Chun-I Lu and Ziqi Sun and Yu-Chen Chang and Wen-Hao Chang and Ye-Ru Chen and Akiyoshi Park and Beining Rao and Siyuan Qiu and Yann-Wen Lan and Ting-Hua Lu and Nai-Chang Yeh},
  title   = {Cryogenic scanning photocurrent spectroscopy for materials responses to structured optical fields},
  journal = {Rev. Sci. Instrum.},
  volume  = {97},
  pages   = {013906},
  year    = {2026},
  doi     = {10.1063/5.0272829}
}

@article{Sato2026npj,
  author  = {Takumi Sato and Satoru Hayami},
  title   = {Quantum theory of magnetic octupole in periodic crystals and application to {$d$}-wave altermagnets},
  journal = {npj Quantum Mater.},
  volume  = {11},
  pages   = {32},
  year    = {2026},
  doi     = {10.1038/s41535-026-00865-9}
}

@article{Maiani2026PRL,
  author  = {Maiani, Andrea},
  title   = {Emergent gauge fields and linear dichroism in spin-textured altermagnets},
  journal = {Phys. Rev. Lett.},
  year    = {2026},
  note    = {accepted 24 August 2026},
  doi     = {10.1103/phr7-1zy5}
}

@article{Ezawa2026PRB,
  author  = {Ezawa, Motohiko},
  title   = {Spin-selective elliptic optical dichroism and perfectly spin-polarized third-order nonlinear photocurrent in altermagnets},
  journal = {Phys. Rev. B},
  volume  = {114},
  pages   = {L111201},
  year    = {2026},
  doi     = {10.1103/5rrq-5g1k}
}

@misc{Ghorashi2026arXiv,
  author        = {Ghorashi, Sayed Ali Akbar and Rappe, Andrew M.},
  title         = {Multiphoton fingerprints of altermagnetic spin splittings},
  year          = {2026},
  eprint        = {2606.27548},
  archivePrefix = {arXiv},
  primaryClass  = {cond-mat.mes-hall},
  doi           = {10.48550/arXiv.2606.27548}
}

@article{Yuan2026PRL,
  author  = {Yuan, Rundong and Jankowski, Wojciech J. and Shen, Ka and Slager, Robert-Jan},
  title   = {Quantum geometry of altermagnetic magnons probed by light},
  journal = {Phys. Rev. Lett.},
  year    = {2026},
  note    = {accepted 5 August 2026},
  doi     = {10.1103/12cl-b9jj}
}

@misc{Sivianes2026arXiv,
  author        = {Sivianes, Javier and Boquete-Someso, Enrique and Hernang{\'o}mez-P{\'e}rez, Daniel and Iba{\~n}ez-Azpiroz, Julen},
  title         = {Optical response beyond magnetic symmetries},
  year          = {2026},
  eprint        = {2608.16368},
  archivePrefix = {arXiv},
  primaryClass  = {cond-mat.mtrl-sci},
  doi           = {10.48550/arXiv.2608.16368}
}

@article{Takahashi2025npj,
  author  = {Takahashi, Koichiro and Huang, Hong-Fei and Yu, Jie-Xiang and Zang, Jiadong},
  title   = {Symmetry and minimal Hamiltonian of nonsymmorphic collinear antiferromagnet {MnTe}},
  journal = {npj Quantum Mater.},
  volume  = {10},
  pages   = {70},
  year    = {2025},
  doi     = {10.1038/s41535-025-00784-1}
}

@article{Guo2025LSA,
  author  = {Guo, Xin and Wang, Pan and Lyu, Guowei and others},
  title   = {Weak-disturbance imaging and characterization of ultra-confined optical near fields},
  journal = {Light Sci. Appl.},
  volume  = {14},
  pages   = {358},
  year    = {2025},
  doi     = {10.1038/s41377-025-01951-6}
}

@article{deLaFiguera2017,
  author  = {de la Figuera, Juan and Tusche, Christian},
  title   = {The Verwey transition observed by spin-resolved photoemission electron microscopy},
  journal = {Appl. Surf. Sci.},
  volume  = {391},
  pages   = {66--69},
  year    = {2017},
  doi     = {10.1016/j.apsusc.2016.05.140}
}

@article{Kresse1996VASP,
  author  = {Kresse, Georg and Furthmuller, Jurgen},
  title   = {Efficient iterative schemes for ab initio total-energy calculations using a plane-wave basis set},
  journal = {Phys. Rev. B},
  volume  = {54},
  pages   = {11169--11186},
  year    = {1996},
  doi     = {10.1103/PhysRevB.54.11169}
}

@article{autieri2026relativistic,
  title={Relativistic spin-momentum locking in altermagnets},
  author={Autieri, Carmine and Fakhredine, Amar},
  journal={The Journal of Physical Chemistry Letters},
  volume={17},
  number={2},
  pages={449--455},
  year={2026},
  publisher={ACS Publications}
}

@article{Kresse1999VASP,
  author  = {Kresse, Georg and Joubert, Daniel},
  title   = {From ultrasoft pseudopotentials to the projector augmented-wave method},
  journal = {Phys. Rev. B},
  volume  = {59},
  pages   = {1758--1775},
  year    = {1999},
  doi     = {10.1103/PhysRevB.59.1758}
}

@article{Blochl1994PAW,
  author  = {Bl{\"o}chl, Peter E.},
  title   = {Projector augmented-wave method},
  journal = {Phys. Rev. B},
  volume  = {50},
  pages   = {17953--17979},
  year    = {1994},
  doi     = {10.1103/PhysRevB.50.17953}
}

@article{PBE1996,
  author  = {Perdew, John P. and Burke, Kieron and Ernzerhof, Matthias},
  title   = {Generalized gradient approximation made simple},
  journal = {Phys. Rev. Lett.},
  volume  = {77},
  pages   = {3865--3868},
  year    = {1996},
  doi     = {10.1103/PhysRevLett.77.3865}
}

@article{Liechtenstein1995,
  author  = {Liechtenstein, A. I. and Anisimov, V. I. and Zaanen, J.},
  title   = {Density-functional theory and strong interactions: Orbital ordering in Mott-Hubbard insulators},
  journal = {Phys. Rev. B},
  volume  = {52},
  pages   = {R5467--R5470},
  year    = {1995},
  doi     = {10.1103/PhysRevB.52.R5467}
}

@article{Monkhorst1976,
  author  = {Monkhorst, Hendrik J. and Pack, James D.},
  title   = {Special points for Brillouin-zone integrations},
  journal = {Phys. Rev. B},
  volume  = {13},
  pages   = {5188--5192},
  year    = {1976},
  doi     = {10.1103/PhysRevB.13.5188}
}

@misc{SM,
  author = {{Supplemental Material}},
  title  = {Supplemental Material for ``Rank-Selective Optical Tomography of Higher-Wave Altermagnetism''},
  year   = {2026},
  note   = {See Supplemental Material at [URL will be inserted by the publisher] for derivations of the rank-selection and aliasing criteria, finite-momentum response formulas, numerical convergence, experimental sampling protocols, and first-principles details.}
}

\end{document}


\title{Supplemental Material for ``Rank-Selective Optical Tomography of Higher-Wave Altermagnetism''}
\author{Meysam Bagheri Tagani}
\author{Carmine Autieri}
\author{Sahar Izadi Vishkayi}
\date{\today}
\maketitle

This Supplemental Material provides the finite-momentum response theory,
proofs of the rotational and crystalline selection rules, microscopic
implementation and convergence tests, experimental sampling analysis, and
first-principles details for MnTe.  It separates the continuum rank-matching
statement from the exact crystal-group criterion and from the material-specific
dynamics.

\section{Rank-selective optical spatial dispersion.}

\subsection{Higher-wave altermagnetic order and multipole conventions}

We consider a collinear altermagnet whose non-relativistic spin splitting is even in
crystal momentum.  Choosing a spin axis $a$, the leading pure $\ell$-wave component
may be written locally in momentum space as
\begin{equation}
\Delta_a(\bm k)=
\Lambda_{a\langle i_1\cdots i_\ell\rangle}
 k_{i_1}\cdots k_{i_\ell}+\cdots,
\qquad \ell=2,4,6,\ldots,
\label{S1}
\end{equation}
where angular brackets denote the symmetric-traceless (ST) part in the spatial
indices.  The real-space magnetic multipole with the same spatial symmetry is
\begin{equation}
\mathcal M_{a\langle i_1\cdots i_\ell\rangle}
=\int_{\rm cell} d^d r\,\mu_a(\bm r)\,
 r_{\langle i_1}\cdots r_{i_\ell\rangle}.
\label{S2}
\end{equation}
For even $\ell$, this object is inversion-even and time-reversal odd.  The mapping
between the momentum-space harmonic and the real-space magnetic multipole is the
standard multipolar description of higher-wave altermagnetism
\cite{Bhowal2024PRX,McClarty2024PRL,Sunko2026npj}; a gauge-invariant quantum
formulation of the octupole in periodic crystals was recently given in
Ref.~\cite{Sato2026npj}.

For two dimensions it is useful to introduce $k_{\pm}=k_x\pm i k_y$.  The two real
basis functions of a pure $\ell$ harmonic are the real and imaginary parts of
$k_+^\ell$:
\begin{align}
\ell=2:\quad &
\Phi_2^{c}=k_x^2-k_y^2,\qquad
\Phi_2^{s}=2k_xk_y,\\
\ell=4:\quad &
\Phi_4^{c}=k_x^4-6k_x^2k_y^2+k_y^4,\nonumber\\
&\Phi_4^{s}=4k_xk_y(k_x^2-k_y^2),\\
\ell=6:\quad &
\Phi_6^{c}=k_x^6-15k_x^4k_y^2+15k_x^2k_y^4-k_y^6,\nonumber\\
&\Phi_6^{s}=6k_x^5k_y-20k_x^3k_y^3+6k_xk_y^5.
\end{align}

The planar basis functions above are representative continuum harmonics used
to illustrate the rank classification and should not be identified with every
material-specific crystalline basis function.  In three dimensions, the same
spatial rank $\ell$ can have a different in-plane angular winding.  For
example, the nonrelativistic bulk $g$-wave order of hexagonal MnTe belongs to
a rank-four sector represented, in continuum notation, by
\[
\Phi_{4}^{\rm MnTe}(\bm k)
\propto
k_z k_y(3k_x^2-k_y^2)
=
k_z k_\parallel^3\sin 3\phi_k ,
\]
up to the crystallographic-axis convention.  Hence the $m=3$ in-plane
winding does not imply $\ell=3$: the additional out-of-plane momentum factor
makes the full spatial polynomial fourth order.

For the local analysis around the $A$ point used in our first-principles
calculations, periodicity of the crystal should be kept in mind.  The DFT
texture is locally consistent with
\[
\Phi_{4,A}^{\rm MnTe}
\propto
\delta k_z\,k_\parallel^3\sin 3\phi_k,
\qquad
\delta k_z=k_z-k_A,
\]
as evidenced by the $m=3$ angular harmonic on fixed-$k_z$ sections and the
reversal of its signed Fourier coefficient on opposite sides of $A$.
Thus the observed fixed-plane $m=3$ winding together with the odd
out-of-plane dependence realizes a three-dimensional $\ell=4$ texture.

The continuum rank-matching statement itself is independent of this choice
of basis component: a rank-four magnetic tensor can couple to the rank-two
polarization tensor through two powers of probe momentum,
$\mathcal M_{a\langle ijkl\rangle}P_{ij}q_kq_l$, giving the characteristic
$q^2$ spatial-dispersion order whenever the corresponding tensor contraction
is allowed by the probe geometry.  The detailed polarization and
momentum-angle projector, however, is material dependent.  In particular,
for a three-dimensional component carrying a $z$ index, that index must be
supplied by the polarization tensor or by the photon momentum.

Thus $d$-, $g$-, and $i$-wave textures correspond to an axial spin index
multiplied by spatial tensors of ranks two, four, and six, respectively; the
associated magnetic multipoles have total ranks three, five, and seven.
For the planar continuum sector used below, we introduce the shorthand
complex amplitude
\begin{equation}
\eta_\ell=|\eta_\ell|e^{i\ell\phi_{\mathcal M}},
\label{S6}
\end{equation}
where $\phi_{\mathcal M}$ specifies the orientation of that planar continuum
harmonic.

\subsection{Exact finite-momentum spin-resolved absorption}

The theory is formulated at finite photon momentum before any multipole expansion.
For a monochromatic vector potential
\begin{equation}
\bm A(\bm r,t)=\bm A_0 e^{i(\bm q\cdot\bm r-\omega t)}+\mathrm{c.c.},
\qquad \bm E_0=i\omega\bm A_0,
\label{S7}
\end{equation}
the Fourier component of the current operator is defined by the functional
derivative
\begin{equation}
\hat j_i(\bm q)=-\left.
\frac{\delta \hat H[\bm A]}{\delta A_i(-\bm q)}\right|_{\bm A=0}.
\label{S8}
\end{equation}
This definition is preferable to replacing the finite-$q$ vertex by
$\partial_{k_i}H(\bm k)$, which is valid only in the local electric-dipole limit.
It also makes contact with gauge-covariant formulations of optical spatial
dispersion in periodic crystals~\cite{Pozo2023SciPost}.

Let $\bm k_{\pm}=\bm k\pm\bm q/2$ and define the exact interband current vertex
\begin{equation}
J^i_{cv}(\bm k,\bm q)=
\langle u_{c\bm k_+}|\hat j_i(\bm q)|u_{v\bm k_-}\rangle.
\label{S9}
\end{equation}
For a spin-analysis direction $\hat{\bm n}$, introduce conduction-state projectors
$\hat P_{\lambda}^{\bm n}=(1+\lambda\,\hat{\bm n}\cdot\hat{\bm s})/2$ with
$\lambda=\pm$.  For an isolated conduction band, or after neglecting interband coherences between nondegenerate final states, the net transition rate resolved by the spin projector $\hat P_{\lambda}^{\bm n}$ is
\begin{align}
W_{\lambda\hat{\bm n}}(\bm q,\omega,\bm e)
=&\frac{2\pi}{\hbar}\frac{|E_0|^2}{\omega^2}
\sum_{cv\bm k}
F_{cv}(\bm k,\bm q)
\left|e_iJ^i_{cv}(\bm k,\bm q)\right|^2\nonumber\\
&\times
\langle u_{c\bm k_+}|\hat P_{\lambda}^{\bm n}|u_{c\bm k_+}\rangle
\delta\!\left(E_{c\bm k_+}-E_{v\bm k_-}-\hbar\omega\right),
\label{S10}
\end{align}
where
$F_{cv}=f(E_{v\bm k_-})-f(E_{c\bm k_+})$.
Equation~(\ref{S10}) is intended as the independent-particle form used later for
microscopic calculations; excitonic corrections can be included by replacing the
single-particle transitions by excitonic states without changing the symmetry
analysis below.

The spin contrast is
\begin{equation}
\Delta W_{\hat{\bm n}}=W_{+\hat{\bm n}}-W_{-\hat{\bm n}}
=n_a e_i^*\Xi_{a;ij}(\bm q,\omega)e_j.
\label{S11}
\end{equation}
In the spin-conserving non-relativistic limit, this is exactly the difference of the
two spin-channel absorption rates.  With spin-orbit coupling, Eq.~(\ref{S10})
provides a continuous spin-projection weight instead of an exact $\uparrow/\downarrow$
label.

To isolate the magnetic contribution from structural and other domain-even
backgrounds, we antisymmetrize the response under reversal of the
altermagnetic domain.  Denoting the set of local magnetic moments in one
domain by $\{\bm m_\alpha\}$, the reversed domain is obtained by
$\bm m_\alpha\rightarrow-\bm m_\alpha$ for every magnetic site $\alpha$,
while keeping the atomic structure fixed.  We therefore define
\begin{equation}
\Xi^{\rm odd}_{a;ij}(\bm q,\omega)
=
\frac{1}{2}
\left[
\Xi_{a;ij}\!\left(\bm q,\omega;\{\bm m_\alpha\}\right)
-
\Xi_{a;ij}\!\left(\bm q,\omega;\{-\bm m_\alpha\}\right)
\right].
\label{S12}
\end{equation}
Any contribution that is unchanged by magnetic-domain reversal, such as
ordinary structural anisotropy, cancels in Eq.~(\ref{S12}), whereas terms
odd in the altermagnetic order are retained.  For a pure $\ell$-wave state,
the same operation reverses the corresponding time-reversal-odd magnetic
multipole,
$\mathcal M_{a\langle i_1\cdots i_\ell\rangle}
\rightarrow
-\mathcal M_{a\langle i_1\cdots i_\ell\rangle}$.

For linear dichroism, only the symmetric-traceless polarization sector is needed. In
two dimensions,
\begin{equation}
P_{ij}=\frac{e_i^*e_j+e_j^*e_i}{2}
-\frac{\delta_{ij}}{2}|\bm e|^2,
\label{S13}
\end{equation}
which transforms as angular momentum $m=\pm2$.  Circular-polarization terms reside
in the antisymmetric sector and are not part of the rank hierarchy discussed here.

\subsection{Rotational rank-matching theorem}

We now prove the continuum result used in the main text.  Introduce
\begin{equation}
q_\pm=q_x\pm iq_y,
\qquad
P_\pm=(P_{xx}-P_{yy})\pm2iP_{xy}.
\label{S14}
\end{equation}
Under an in-plane rotation by $\alpha$,
\begin{equation}
q_+\rightarrow e^{i\alpha}q_+,
\qquad
P_+\rightarrow e^{2i\alpha}P_+,
\qquad
\eta_\ell\rightarrow e^{i\ell\alpha}\eta_\ell.
\label{S15}
\end{equation}
It is convenient to express the in-plane photon momentum in the circular
basis,
$q_{\pm}=q_x\pm iq_y=q\,e^{\pm i\phi_q}$.
Under an in-plane rotation by $\alpha$,
$q_{\pm}\rightarrow e^{\pm i\alpha}q_{\pm}$, so the two components carry
angular harmonics $m_q=\pm1$.  A homogeneous polynomial of degree $n$ is
spanned by monomials $q_+^r q_-^{n-r}$, which transform as
$e^{i(2r-n)\alpha}$.  Hence the $n$th symmetric product of the in-plane
vector representation contains the angular harmonics
\begin{equation}
m_q=n,n-2,\ldots,-n.
\label{S16}
\end{equation}
In particular, $|m_q|\le n$ and $m_q$ has the same parity as $n$.
Therefore, a scalar linear in $\eta_\ell$, linear in the polarization tensor, and of
order $q^n$ requires angular-momentum conservation
\begin{equation}
\sigma_{\mathcal M}\ell+\sigma_P 2+m_q=0,
\qquad \sigma_{\mathcal M},\sigma_P=\pm1.
\label{S17}
\end{equation}
For $\ell\ge2$ the smallest possible $n$ is obtained with
$m_q=\ell-2$, giving
\begin{equation}
n_{\min}^{O(2)}=\ell-2.
\label{S18}
\end{equation}
No lower order can contribute to the pure $\ell$ harmonic because a degree-$n$
homogeneous polynomial in $\bm q$ cannot contain an angular Fourier component with
$|m_q|>n$.  At the threshold order, the invariant is unique up to its coefficient
and complex conjugation,
\begin{equation}
\mathcal I_\ell=
\eta_\ell^* P_+q_+^{\ell-2}
+\eta_\ell P_-q_-^{\ell-2}.
\label{S19}
\end{equation}
The Cartesian form of Eq.~(\ref{S19}) is precisely
\begin{equation}
\mathcal I_\ell\propto
n_a\mathcal M_{a\langle i_1\cdots i_\ell\rangle}
P_{i_1i_2}q_{i_3}\cdots q_{i_\ell}.
\label{S20}
\end{equation}
For a centrosymmetric even-parity altermagnet, inversion sends
$\bm q\rightarrow-\bm q$ while leaving the axial spin index and the even-$\ell$
magnetic multipole invariant.  Because $P_{ij}$ is also inversion even, only even
$n$ occur in this domain-odd linear-dichroic sector, consistent with
$n=0,2,4$ for $d$, $g$, and $i$ waves.

For linearly polarized light,
$\bm e=(\cos\theta,\sin\theta)$, and
$\bm q=q(\cos\phi_q,\sin\phi_q)$, Eq.~(\ref{S19}) becomes
\begin{equation}
\Delta W_\ell^{\rm LD}
=\widetilde C_\ell(\omega)|\eta_\ell|
q^{\ell-2}
\cos\!\left[\ell\phi_{\mathcal M}-2\theta-(\ell-2)\phi_q\right]
+\mathcal O(q^\ell).
\label{S21}
\end{equation}
The three lowest cases are summarized in Table~\ref{tab:hierarchy}.

\begin{table*}[h]
\caption{Rotationally resolved hierarchy for linearly polarized light.}
\label{tab:hierarchy}
\begin{ruledtabular}
\begin{tabular}{c c c c}
order & spatial harmonic & leading $q$ power & joint angular harmonic \\
\hline
$d$ ($\ell=2$) & $e^{\pm2i\phi}$ & $q^0$ & $2\theta$ \\
$g$ ($\ell=4$) & $e^{\pm4i\phi}$ & $q^2$ & $2\theta+2\phi_q$ \\
$i$ ($\ell=6$) & $e^{\pm6i\phi}$ & $q^4$ & $2\theta+4\phi_q$ \\
\end{tabular}
\end{ruledtabular}
\end{table*}

\subsection{Angular tomography and immunity to lower orders in $q$}

The rank-resolved component is extracted by the joint Fourier projection
\begin{equation}
\mathcal T_\ell(q,\omega)=
\int_0^{2\pi}\frac{d\theta}{2\pi}
\int_0^{2\pi}\frac{d\phi_q}{2\pi}
 e^{+i[2\theta+(\ell-2)\phi_q]}
\Delta W_{\hat{\bm n}}(q,\omega;\theta,\phi_q).
\label{S22}
\end{equation}
To see why lower spatial orders cannot contaminate this projection, expand an
analytic response as
\begin{equation}
\Delta W(q,\theta,\phi_q)
=\sum_{n=0}^{\infty}q^n
\sum_{m=-n,-n+2,\ldots,n}
\sum_{p=0,\pm2}
C_{nmp}\,e^{im\phi_q}e^{ip\theta}.
\label{S23}
\end{equation}
The restriction $|m|\le n$ follows solely from polynomial degree and is independent
of crystal symmetry.  Equation~(\ref{S22}) selects the conjugate Fourier sector
$p=-2$ and $m=-(\ell-2)$ in the convention of Eq.~(\ref{S23}); equivalently one may project the complex-conjugate sector.  Hence every term with $n<\ell-2$ vanishes identically.  For the pure
higher-wave invariant,
\begin{equation}
\mathcal T_\ell(q,\omega)=
\mathcal C_\ell(\omega)\eta_\ell q^{\ell-2}
+\mathcal O(q^\ell).
\label{S24}
\end{equation}
Thus the exponent of $q$ and the phase of $\mathcal T_\ell$ provide independent
information: the former gives the spatial rank of the selected channel and the
latter tracks the orientation of the corresponding magnetic harmonic.  The
projection does not assert that the total response lacks lower-$q$ terms.

Since $\Delta W_{\hat{\bm n}}$ has already been defined as the part of the
spin-resolved absorption contrast that is odd under the reversal of all local magnetic moments,
$\{\bm m_\alpha\}\rightarrow\{-\bm m_\alpha\}$,
the tomographic coefficient $\mathcal T_\ell$ obtained from
Eq.~(\ref{S22}) is itself domain odd. 

For centrosymmetric samples, it is useful to further symmetrize the
domain-odd response under reversal of the photon momentum,
\begin{equation}
\Delta W_{\hat{\bm n}}^{\,q{\rm -even}}(\bm q)
=
\frac{1}{2}
\left[
\Delta W_{\hat{\bm n}}(\bm q)
+
\Delta W_{\hat{\bm n}}(-\bm q)
\right],
\label{S25}
\end{equation}
which removes contributions that are odd in $\bm q$, such as conventional
photon-drag or other nonreciprocal spatial-dispersion backgrounds, while
retaining the even-$q$ rank-resolved channels
$q^0,q^2,q^4,\ldots$ relevant to even-parity $d$-, $g$-, and $i$-wave
altermagnetism.

\subsection{Discrete point groups and crystalline aliasing}

The continuum result must be separated from the exact selection rule of a crystal.
Let $G$ be the relevant parent point group (or magnetic point group when SOC is
retained).  Denote by $\Gamma_{\mathcal M}$ the irrep of the altermagnetic order,
$\Gamma_n$ that of the analyzed axial spin component, $\Gamma_P$ that of the
symmetric-traceless linear-polarization tensor, and $\Gamma_q$ the in-plane vector
representation.  A scalar contribution at order $q^n$ exists if and only if
\begin{equation}
A_1\subset
\Gamma_{\mathcal M}\otimes\Gamma_n\otimes\Gamma_P\otimes
\operatorname{Sym}^{n}(\Gamma_q).
\label{S27}
\end{equation}
Consequently,
\begin{equation}
n_{\min}^{\rm crys}=\min\{n\ge0:\ \text{Eq.~(\ref{S27}) holds}\},
\label{S28}
\end{equation}
with inversion parity and mirror character imposed simultaneously.  We call the
case $n_{\min}^{\rm crys}<\ell-2$ crystalline aliasing.  It is not a failure of the
higher-wave classification: it reflects the fact that distinct continuum angular
momenta may subduce to the same irrep of a finite point group.

For a pure cyclic rotation $C_N$, before mirror constraints are imposed, the
condition can be written in angular-momentum language as
\begin{equation}
\sigma_{\mathcal M}\ell+\sigma_P2+m_q=0\pmod N,
\qquad
m_q\in\{n,n-2,\ldots,-n\}.
\label{S29}
\end{equation}
This makes the origin of aliasing transparent.  For example, at $q=0$ a $g$-wave
harmonic ($\ell=4$) and the linear-polarization harmonic ($|m|=2$) can become
rotation-compatible in a sixfold environment because $4+2=0\pmod6$; whether the
term actually survives is then decided by the remaining mirrors and by the spin
representation.  Hence a local optical spin signature of a $g$-wave altermagnet is
not forbidden in general, consistent with recent calculations for
$\alpha$-MnTe~\cite{Haag2026PRM}.  The finite-$q$ tomographic coefficient
$\mathcal T_4$, however, specifically selects the $m_q=2$ channel and therefore
starts at $q^2$ by construction.

The same logic shows why the total crystalline response cannot be inferred
from the labels $d$, $g$, or $i$ alone.  Equation~(\ref{S27}) is the exact
material-specific criterion, whereas $n=\ell-2$ applies to the continuously
rank-resolved Fourier component.

\subsection{Relation to existing optical multipole probes}

The present observable differs from two neighboring strategies in a way that is
important for the symmetry counting.  In linear magneto-birefringence, the measured
dielectric tensor has two polar indices and a static magnetic field supplies the
required time-reversal-odd axial index.  A $d$-wave octupole can therefore couple to
$E_jE_kH_a$, whereas a $g$-wave triakontadipole requires two additional spatial
indices, supplied in linear optics by strain through terms of the form
$E_jE_k\eta_{lm}H_a$~\cite{Sunko2026npj}.  In the present construction, no magnetic
field is used: the measured spin channel supplies the axial index, and
$q_lq_m$ supplies the two additional spatial indices for the $g$-wave response.
Likewise, four powers of $q$ supply the additional spatial rank for an $i$-wave
component.

Gradient-assisted second-harmonic generation provides a different route in which
spatial gradients enter a nonlinear polarization.  The recent octupolar SHG
experiment in CoF$_2$ identified an electric-quadrupolar contribution of the form
$\bm P^{2\omega}\sim\mathcal O^M:\bm E^\omega\nabla\bm E^\omega$
\cite{Usachev2026PRL}.  Our response remains a one-photon, linear-response
absorption process; spatial dispersion is used to resolve the tensor rank of the
spin-selective transition probability rather than to generate a harmonic at
$2\omega$.  Finally, the nonlinear spin-current hierarchy proposed for $g$- and $i$-wave altermagnets increases the order in the electric field
\cite{Ezawa2025PRB}, whereas the present hierarchy keeps the optical field order
fixed and increases the order in spatial momentum.

\subsection{Robustness to weak spin-orbit coupling}

The rank argument is clearest in the non-relativistic spin-group limit, where the two
spin sectors are exactly conserved.  Weak spin-orbit coupling does not invalidate
the definition of Eq.~(\ref{S10}); it replaces the binary spin label by the
expectation value of the projector $\hat P_{\lambda}^{\bm n}$.  The exact symmetry criterion must then be evaluated in the magnetic point
group rather than in the direct product of spin and spatial groups; complementary
spin-space-group constraints on optical tensors have recently been developed in
Ref.~\cite{Sivianes2026arXiv}.  Two consequences should be kept
separate.

First, SOC can activate additional lower-rank domain-odd optical tensors that are
forbidden in the non-relativistic limit.  Such terms are material-dependent and are
part of the crystalline-aliasing/background sector.  Second, SOC cannot make a
homogeneous polynomial of degree $n$ in $\bm q$ acquire a momentum-angle Fourier
harmonic with $|m_q|>n$.  Therefore the kinematic statement underlying
Eq.~(\ref{S24}) survives: the $(\ell-2)$th momentum-angle harmonic cannot appear
below order $q^{\ell-2}$.  What SOC can change is the coefficient and the degree to which that harmonic
is uniquely attributable to the non-relativistic higher-wave order.  The
first-principles MnTe analysis below quantifies this distinction directly.

\section{Rank-pure microscopic realization}

\subsection{Rank-pure microscopic benchmark}

To test the rank-selection rule independently of lattice-specific subduction, we
use the continuum Hamiltonian
\begin{equation}
H_{\ell}(\bm k)
=
\hbar v\left(k_x\rho_z\tau_x+k_y\tau_y\right)
+
\left[m+\Delta_{\ell}\Phi_{\ell}(\bm k)s_z\right]\tau_z ,
\label{S:model}
\end{equation}
with
\begin{equation}
\Phi_{\ell}(\bm k)
=
k_0^{-\ell}
\mathrm{Re}\!\left[
e^{-i\ell\phi_{\mathcal M}}(k_x+ik_y)^\ell
\right],
\qquad
\ell=2,4,6.
\label{S:harmonic}
\end{equation}
The Pauli matrices $\tau_i$, $s_i$, and $\rho_i$ act in orbital, physical-spin,
and time-reversed orbital-flavor spaces.  The nonmagnetic parent Hamiltonian
$H_0$ is invariant under
\begin{equation}
\mathcal P=\tau_z,
\qquad
\mathcal T=is_y\rho_xK,
\qquad
\mathcal T^2=-1 ,
\label{S:PT}
\end{equation}
and, because $H_0$ is spin independent, has exact spin $SU(2)$ symmetry.
For even $\ell$, $\Phi_\ell(-\bm k)=\Phi_\ell(\bm k)$, so the magnetic term is
inversion even, while $\mathcal T s_z\mathcal T^{-1}=-s_z$ makes it
time-reversal odd.  The spectrum is independent of $\rho$,
\begin{equation}
E_{\lambda s}(\bm k)
=
\lambda\sqrt{
(\hbar vk)^2+
\left[m+s\Delta_\ell\Phi_\ell(\bm k)\right]^2},
\label{S:spectrum}
\end{equation}
and obeys
$E_{\lambda,+}(\bm k)
=
E_{\lambda,-}(R_{\pi/\ell}\bm k)$ because
$\Phi_\ell(R_{\pi/\ell}\bm k)=-\Phi_\ell(\bm k)$.  The model is therefore a
rank-pure non-relativistic benchmark; lower-rank crystalline aliasing is
intentionally absent.

\subsection{Ward-consistent finite-\texorpdfstring{$q$}{q} current vertex}

Let $\bm k_\pm=\bm k\pm\bm q/2$.  For each spin and orbital-flavor block
$h_{s\rho}(\bm k)$ we define
\begin{equation}
\mathcal V_i^{(s\rho)}(\bm k,\bm q)
=
\frac{1}{\hbar}
\int_{-1/2}^{1/2}d\lambda\,
\partial_{k_i}
h_{s\rho}(\bm k+\lambda\bm q).
\label{S:vertex}
\end{equation}
This line-averaged vertex has two useful properties.  First,
$\mathcal V_i(\bm k,\bm 0)
=
\hbar^{-1}\partial_{k_i}h(\bm k)$.
Second, contraction with photon momentum gives
\begin{align}
q_i\mathcal V_i(\bm k,\bm q)
&=
\frac{1}{\hbar}
\int_{-1/2}^{1/2}d\lambda\,
\frac{d}{d\lambda}
h(\bm k+\lambda\bm q)
\nonumber\\
&=
\frac{
h(\bm k_+)-h(\bm k_-)
}{\hbar},
\label{S:ward}
\end{align}
so the longitudinal Ward identity is satisfied exactly within the continuum
model.  Numerically, the integral in Eq.~(\ref{S:vertex}) is evaluated by
six-point Gauss--Legendre quadrature.

For an insulating ground state, the symmetric absorptive tensor for spin
$s=\pm$ is evaluated as
\begin{equation}
S_{ij}^{(s)}(\bm q,\omega)
=
2\sum_{\rho}^{\phantom{\rho}\prime}
\int\frac{d^2k}{(2\pi)^2}
\frac{
\mathrm{Re}\!\left[
J_{i,s}^{*}J_{j,s}
\right]
}{
\omega_{cv,s}
}
\delta_\gamma\!
\left(\omega-\omega_{cv,s}\right),
\label{S:opticaltensor}
\end{equation}
where the prime indicates that one $\rho$ block is evaluated explicitly and
the prefactor $2$ restores the exactly degenerate time-reversed partner,
\begin{align}
\label{S:matrixelement}
J_{i,s}
=
\langle
u_{c,s}(\bm k_+)
|
\mathcal V_i^{(s)}(\bm k,\bm q)
|
u_{v,s}(\bm k_-)
\rangle ,
\\ \nonumber
\omega_{cv,s}
=
\frac{
E_{c,s}(\bm k_+)-E_{v,s}(\bm k_-)
}{\hbar}.
\end{align}
Equation~\eqref{S:opticaltensor} is the symmetric absorptive-conductivity tensor used for the normalized microscopic benchmark. In the zero-broadening limit at fixed external frequency, it differs from the transition-rate tensor of Eq. (10) of the paper only by an overall frequency-dependent factor; consequently the normalized \(q\)-scaling exponents and tomographic phases are unchanged. Overall electromagnetic prefactors cancel from the normalized comparisons
used here and are omitted.  We use the spin-odd tensor
\begin{equation}
\Delta S_{ij}=S_{ij}^{(+)}-S_{ij}^{(-)}.
\label{S:spinodd}
\end{equation}
For this spin-conserving model, reversing the altermagnetic domain exchanges
the two spin sectors, so Eq.~(\ref{S:spinodd}) is already domain odd.

\subsection{Direct extraction of the tomographic harmonic}

For a linearly polarized field
$\bm e=(\cos\theta,\sin\theta)$,
\begin{equation}
\Delta W(\theta,\phi_q)
=
e_i\Delta S_{ij}(\phi_q)e_j
=
A_0+A_c\cos2\theta+A_s\sin2\theta,
\label{S:thetaexpansion}
\end{equation}
with
\begin{equation}
A_c=\frac{\Delta S_{xx}-\Delta S_{yy}}{2},
\qquad
A_s=\Delta S_{xy}.
\end{equation}
The $\theta$ integration appearing in the main-text tomography can therefore
be carried out analytically:
\begin{equation}
F_2(\phi_q)
\equiv
\int_0^{2\pi}\frac{d\theta}{2\pi}
e^{2i\theta}\Delta W
=
\frac{
\Delta S_{xx}-\Delta S_{yy}
}{4}
+
\frac{i}{2}\Delta S_{xy}.
\label{S:F2}
\end{equation}
The projected coefficient then reduces to the one-dimensional angular
integral
\begin{equation}
\mathcal T_\ell(q,\omega)
=
\int_0^{2\pi}\frac{d\phi_q}{2\pi}
e^{i(\ell-2)\phi_q}
F_2(\phi_q).
\label{S:Tnumeric}
\end{equation}
Equation~(\ref{S:Tnumeric}) is the expression evaluated numerically for
Fig.~2 of the main text.

\subsection{Numerical protocol, convergence, and asymptotic scaling}

We use $E_0=\hbar v k_0$ as the energy unit and $k_0$ as the momentum unit.
The parameters of Fig.~2 are
\begin{equation}
m/E_0=0.60,\qquad
\Delta_\ell/E_0=0.08,\qquad
\hbar\omega/E_0=2.00 .
\end{equation}
The Dirac delta is represented numerically by
\begin{equation}
\delta_\gamma(x)
=
\frac{1}{\sqrt{\pi}\gamma}
e^{-x^2/\gamma^2},
\qquad
\gamma/E_0=0.04.
\label{S:gaussian}
\end{equation}
The radial cutoff is $k_{\rm max}=1.4k_0$.  The production grid contains
$100$ radial points, $192$ momentum angles, and $48$ photon-momentum angles.
The asymptotic fit uses
$0.006\le q/k_0\le0.045$.

Table~\ref{tab:conv} shows that the fitted powers are converged well beyond the
precision needed to distinguish the three channels.
\begin{table}[h]
\caption{Convergence of the fitted exponent
$|\mathcal T_\ell|\propto q^{p_\ell}$ over the main-text fitting window.}
\label{tab:conv}
\begin{ruledtabular}
\begin{tabular}{cccccc}
$N_r$ & $N_{\phi_k}$ & $N_{\phi_q}$ & $p_2$ & $p_4$ & $p_6$\\
\hline
60  & 120 & 32 & $-1.56\times10^{-5}$ & 1.99948 & 3.99811\\
80  & 160 & 40 & $-1.72\times10^{-5}$ & 1.99949 & 3.99811\\
100 & 192 & 48 & $-1.72\times10^{-5}$ & 1.99949 & 3.99811
\end{tabular}
\end{ruledtabular}
\end{table}

Changing the radial cutoff from $1.2k_0$ to $2.0k_0$ changes the fitted
exponents by less than $5\times10^{-4}$ for the Gaussian representation of
the delta function.  Replacing Eq.~(\ref{S:gaussian}) by a Lorentzian
broadening also preserves the hierarchy; for a representative converged grid
we obtain $p_2=2.8\times10^{-5}$, $p_4=2.002$, and $p_6=4.008$.
The spectral amplitudes themselves are nonuniversal and depend on broadening,
frequency, and band parameters.

\begin{figure}[t]
\centering
\includegraphics[width=\columnwidth]{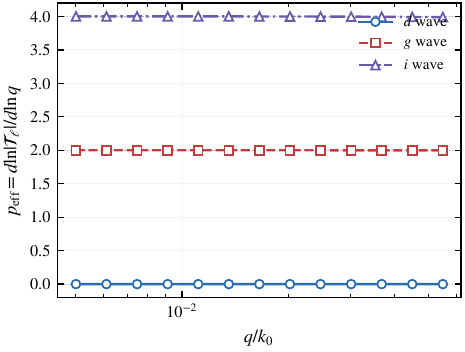}
\caption{\label{fig:Slocal}
Local effective power
$p_{\rm eff}=d\ln|\mathcal T_\ell|/d\ln q$ for the microscopic benchmark.
As $q\rightarrow0$, the three channels approach $0$, $2$, and $4$.
The weak departure of the $i$-wave curve at larger $q$ is the expected
higher-order correction to the gradient expansion, not a change of the
asymptotic selection rule.}
\end{figure}

Figure~\ref{fig:Slocal} makes the asymptotic character explicit.  In
particular, the $i$-wave response gradually departs from the $q^4$ law once
$q$ is no longer small compared with $k_0$, while its local exponent tends
to $4$ in the long-wavelength limit.

\subsection{Fourier purity and domain-phase winding}

\begin{figure}[t]
\centering
\includegraphics[width=\columnwidth]{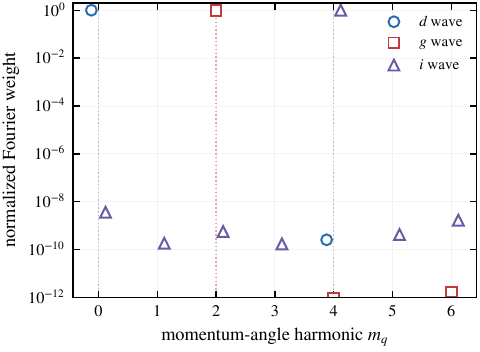}
\caption{\label{fig:Sfourier}
Normalized momentum-angle Fourier spectrum of the linear-polarization
component $F_2(\phi_q)$ at $q=0.03k_0$.  The $d$-, $g$-, and $i$-wave models
select $m_q=0$, $2$, and $4$, respectively, directly realizing the
rank-matching rule.}
\end{figure}

The tomography can also be checked before applying the final
$\phi_q$ projection.  We Fourier analyze $F_2(\phi_q)$ according to
\begin{equation}
F_{2,m}
=
\int_0^{2\pi}\frac{d\phi_q}{2\pi}
e^{im\phi_q}F_2(\phi_q).
\label{S:FourierSpectrum}
\end{equation}
At $q=0.03k_0$, the dominant momentum-angle harmonics are
$m_q=0$, $2$, and $4$ for $\ell=2$, $4$, and $6$, respectively, as shown in
Fig.~\ref{fig:Sfourier}.  Residual weights away from the selected harmonic are
at the numerical integration floor for this rank-pure continuum model.

Finally, rotating the order parameter by $\phi_{\mathcal M}$ multiplies the
selected harmonic by $e^{i\ell\phi_{\mathcal M}}$.  The numerical calculation
therefore gives
\begin{equation}
\arg\mathcal T_\ell(\phi_{\mathcal M})
-
\arg\mathcal T_\ell(0)
=
\ell\phi_{\mathcal M}
\quad (\mathrm{mod}\;2\pi),
\label{S:phasewinding}
\end{equation}
with fitted slopes $2.000000$, $4.000000$, and $6.000000$ for the three
models in Fig.~2(b).  This phase relation is independent of the overall sign
of the nonuniversal spectral coefficient; only phase differences are used to
infer the domain orientation.

\section{Crystalline aliasing and protected tomography}

\subsection{Discrete rotational symmetry and crystalline aliasing}

The continuum rank-selection rule is most transparent in a circular basis.
For an in-plane quantity with angular harmonic $m$, a rotation by $\alpha$
acts as
\begin{equation}
X_m\rightarrow e^{im\alpha}X_m .
\end{equation}
In an $O(2)$-symmetric reference system, an invariant must have zero total
angular harmonic.  In a crystal with only $C_N$ rotational symmetry, angular
harmonics that differ by integer multiples of $N$ transform identically.
Consequently, a contribution of order $q^n$ is rotationally allowed whenever
\begin{equation}
\sigma_{\mathcal M}\ell
+
\sigma_P 2
+
m_q
=
rN,
\qquad
r\in\mathbb Z,
\label{S:CNcriterion}
\end{equation}
for at least one choice
$\sigma_{\mathcal M},\sigma_P=\pm1$, subject also to the mirror, inversion,
and spin-representation constraints of the full point group.  The momentum
harmonic carried by an $n$th-order polynomial satisfies
\begin{equation}
m_q=n,n-2,\ldots,-n .
\label{S:mqset_alias}
\end{equation}
Equation~(\ref{S:CNcriterion}) is the circular-basis form of the representation
criterion given in the main text.

At $q=0$, $m_q=0$.  A lower-order local response can therefore occur if
\begin{equation}
\sigma_{\mathcal M}\ell+\sigma_P2=rN .
\label{S:q0alias}
\end{equation}
For $g$-wave order in a $C_6$ environment, the choice
$\sigma_{\mathcal M}=\sigma_P=+1$ gives
$4+2=6$, so a $q^0$ contribution is permitted even though the
rank-resolved continuum channel begins at $q^2$.  Likewise, for $i$-wave
order in a $C_4$ environment, $6+2=8=2\times4$, allowing a $q^0$
crystalline alias.

The important point is that Eq.~(\ref{S:q0alias}) concerns the
unprojected crystalline response; it does not lower the power required
to generate a specified continuous momentum-angle harmonic.  A homogeneous
polynomial of degree $n$ obeys $|m_q|\le n$.  Therefore the Fourier projection
onto
\begin{equation}
m_q^{\star}=\ell-2
\label{S:targetmq}
\end{equation}
annihilates every contribution with $n<\ell-2$.  This statement is kinematic
and remains valid when continuum angular momenta subduce to the same discrete
crystal representation.  Discrete symmetry can mix the target harmonic with
harmonics differing by multiples of $N$, but it cannot create the actual
continuous Fourier component $m_q^\star$ from a polynomial whose degree is
smaller than $|m_q^\star|$.

\subsection{Hexagonal $g$-wave benchmark}

To realize the aliasing mechanism explicitly, we augment the $g$-wave
Hamiltonian of the main text by a nonmagnetic $C_6$ crystal field,
\begin{equation}
H_{4}^{(C_6)}(\bm k)
=
\hbar v(k_x\rho_z\tau_x+k_y\tau_y)
+
\left[
m+\beta_6\Psi_6(\bm k)
+\Delta_4\Phi_4(\bm k)s_z
\right]\tau_z ,
\label{S:C6model}
\end{equation}
where
\begin{align}
\Phi_4(\bm k)
&=
k_0^{-4}
\mathrm{Re}\!\left[
e^{-4i\phi_{\mathcal M}}
(k_x+ik_y)^4
\right],\\
\Psi_6(\bm k)
&=
k_0^{-6}
\mathrm{Re}\!\left[
e^{-6i\phi_C}
(k_x+ik_y)^6
\right].
\end{align}
The $\beta_6$ term is spin-independent, time-reversal even, inversion even,
and independent of the altermagnetic domain.  It breaks the artificial
continuous rotational symmetry of the reference model while retaining
hexagonal rotational symmetry.  The finite-$q$ velocity vertex is generalized
by replacing the mass derivative in Eq.~(\ref{S:vertex}) with
\begin{equation}
\partial_{k_i}
\left[
\beta_6\Psi_6(\bm k)
+s\Delta_4\Phi_4(\bm k)
\right],
\end{equation}
inside the same straight-line average.  The Ward identity is therefore
preserved exactly.

For linearly polarized light, we first extract
\begin{equation}
F_2(q,\phi_q)
=
\int_0^{2\pi}\frac{d\theta}{2\pi}\,
e^{2i\theta}
\Delta W(q,\theta,\phi_q).
\label{S:F2alias}
\end{equation}
At $q=0$ the photon-momentum direction is immaterial.  The numerical result
shown in Fig.~2(d) of the main text is zero for $\beta_6=0$ and changes linearly with
$\beta_6$ in the weak-anisotropy regime.  Since the response is also odd
under reversal of $\Delta_4$, the leading local alias has the structure
\begin{equation}
F_2(q=0)
=
A_6(\omega)\,
\beta_6\Delta_4
+
\mathcal O(\beta_6^3\Delta_4,\beta_6\Delta_4^3),
\label{S:bilinearalias}
\end{equation}
for the reference orientation used in the numerical calculation.  The
coefficient $A_6$ is nonuniversal.  Equation~(\ref{S:bilinearalias}) makes
clear that the local term is magnetic and domain odd, even though it requires
the domain-even crystal anisotropy to supply the missing angular structure.

\subsection{Fourier protection of the $g$-wave channel}

The target coefficient is
\begin{equation}
\mathcal T_4(q,\omega)
=
\int_0^{2\pi}\frac{d\phi_q}{2\pi}\,
e^{2i\phi_q}F_2(q,\phi_q).
\label{S:T4protected}
\end{equation}
A $q^0$ term is independent of $\phi_q$ and therefore integrates to zero in
Eq.~(\ref{S:T4protected}).  More generally, all terms of order $q^n$ with
$n<2$ are excluded from the $|m_q|=2$ Fourier channel.  For the
centrosymmetric benchmark, inversion additionally eliminates odd powers of
$q$, so the projected expansion has the form
\begin{equation}
\mathcal T_4(q,\omega)
=
\mathcal C_4^{(C_6)}(\omega,\beta_6)\,q^2
+
\mathcal O(q^4).
\label{S:T4C6}
\end{equation}
The coefficient is modified by crystal anisotropy, but the leading power is
not.  Fits over the same long-wavelength regime used in the main text give
\begin{equation}
p_4=
1.999406,\quad
1.999328,\quad
1.999120
\end{equation}
for $\beta_6/E_0=0$, $0.02$, and $0.04$, respectively.

\begin{figure}[t]
\centering
\includegraphics[width=0.48\linewidth]{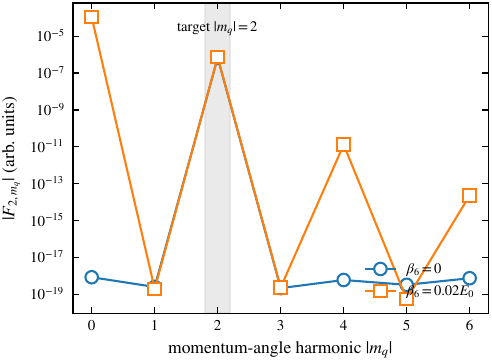}
\caption{\label{fig:S_alias_spectrum}
Momentum-angle Fourier spectrum of the linear-polarization component
$F_2(\phi_q)$ for the $g$-wave benchmark at $q=0.03k_0$.
Hexagonal anisotropy creates a large momentum-local $m_q=0$ contribution,
whereas the target $|m_q|=2$ component remains spectrally distinct and is
selected by the tomographic projection.}
\end{figure}

The Fourier spectrum in Fig.~\ref{fig:S_alias_spectrum} provides a direct diagnostic before
the final projection.  At $\beta_6=0$, the rank-pure model is dominated by the
target $|m_q|=2$ component.  Finite $\beta_6$ produces a much larger
$m_q=0$ component, but this contribution is exactly orthogonal to
Eq.~(\ref{S:T4protected}).  The experiment therefore need not require the
absence of local crystalline signals; it requires sufficient angular control
of $\bm q$ to resolve the selected Fourier harmonic.

\subsection{$i$-wave generalization in a $C_4$ environment}

\begin{figure*}[t]
    \centering
    \includegraphics[width=\textwidth]{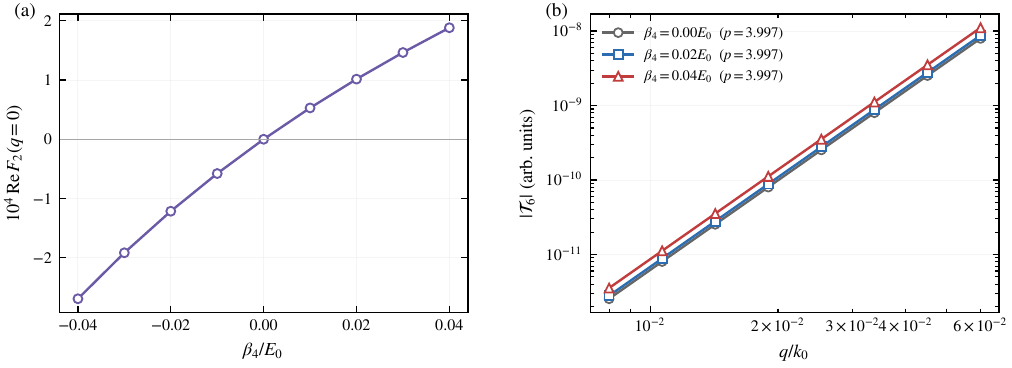}
    \caption{\label{fig:S3alias}
    Crystalline aliasing for $i$-wave order in a $C_4$ environment.
    (a) The local polarization harmonic $F_2(q=0)$ becomes finite when the
    nonmagnetic $C_4$ anisotropy $\beta_4$ is introduced, consistent with the
    discrete-symmetry condition $6+2=2\times4$.
    (b) The rank-projected coefficient $|\mathcal T_6|$ retains its
    $q^4$ long-wavelength scaling for all displayed $\beta_4$, demonstrating
    that the $|m_q|=4$ Fourier channel is protected against the lower-order
    crystalline alias.  Other parameters are those of the microscopic
    benchmark in the main text.}
\end{figure*}

To demonstrate that the protection is not specific to the $g$-wave/$C_6$
example, we consider
\begin{equation}
H_{6}^{(C_4)}(\bm k)
=
H_6(\bm k)
+
\beta_4\Psi_4(\bm k)\tau_z,
\qquad
\Psi_4(\bm k)
=
k_0^{-4}\mathrm{Re}\!\left[(k_x+ik_y)^4\right].
\label{S:C4imodel}
\end{equation}
Here $6+2=8=2\times4$, so $C_4$ symmetry permits a $q^0$ crystalline alias
for the $i$-wave response.  Figure~\ref{fig:S3alias} shows that the local
$F_2(q=0)$ indeed becomes finite as $\beta_4$ is introduced.  Nevertheless,
the $i$-wave tomographic projection selects $|m_q|=4$, and the calculated
response remains
\begin{equation}
\mathcal T_6(q,\omega)
=
\mathcal C_6^{(C_4)}(\omega,\beta_4)\,q^4
+
\mathcal O(q^6).
\end{equation}
The fitted exponents are
$p_6=3.99660$, $3.99664$, and $3.99668$ for
$\beta_4/E_0=0$, $0.02$, and $0.04$, respectively.

\section{Experimental momentum window and discrete
tomography}

\begin{figure*}[t]
\centering
\includegraphics[width=\linewidth]{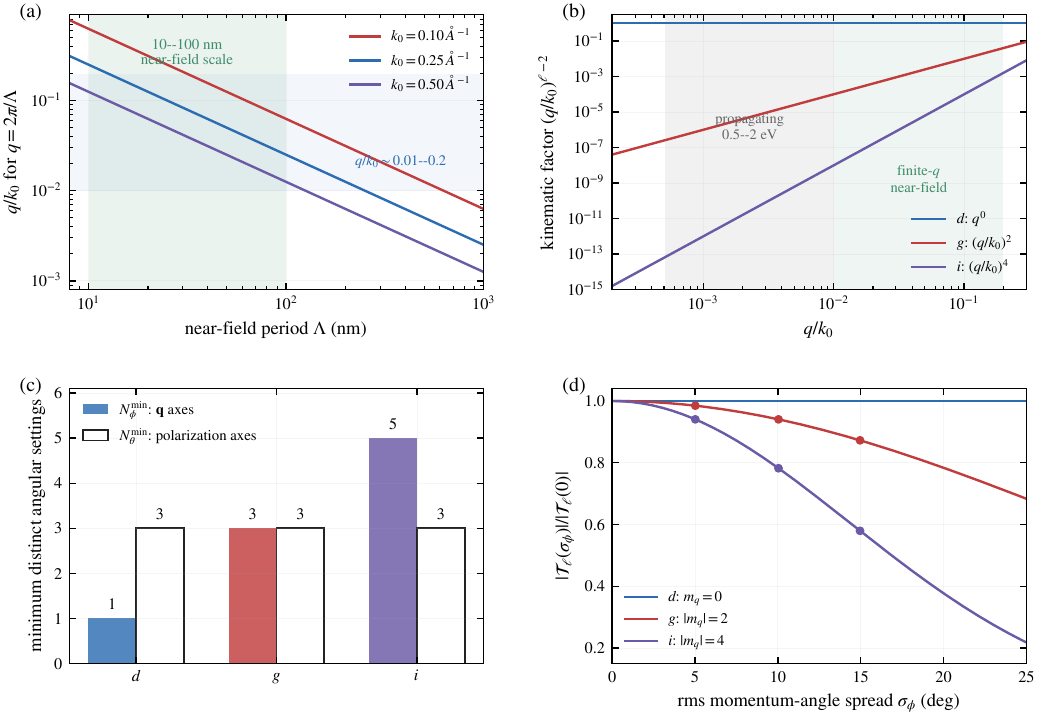}
\caption{\label{fig:S_experiment}
Experimental momentum window and discrete higher-wave tomography.
(a) Dimensionless momentum $q/k_0$ generated by a structured near-field
period $\Lambda$ for representative electronic momentum scales.
(b) Kinematic rank factor $(q/k_0)^{\ell-2}$, with the propagating-light and
near-field windows indicated.
(c) Ideal band-limited minimum number of polarization and momentum axes after
$q$-even symmetrization.
(d) Attenuation caused by a Gaussian rms spread $\sigma_\phi$ of the momentum
direction.}
\end{figure*}

\subsection{Optical momentum scales}

For a propagating mode of angular frequency $\omega$, the in-plane momentum
is bounded by
\begin{equation}
q_{\parallel}\leq q_0=\frac{\omega}{c}.
\label{S:qfree}
\end{equation}
Using $\hbar c=1973.2698~\mathrm{eV\,\angstrom}$,
\begin{equation}
q_0[\angstrom^{-1}]
=
5.0677\times10^{-4}
\left(
\frac{\hbar\omega}{1~\mathrm{eV}}
\right).
\label{S:qfreeunits}
\end{equation}
A structured near field with a well-defined in-plane spatial period
$\Lambda$ supplies the Fourier momentum
\begin{equation}
q=\frac{2\pi}{\Lambda}
=
\frac{0.62832}{\Lambda[\mathrm{nm}]}
~\angstrom^{-1}.
\label{S:qnear}
\end{equation}
Hence $\Lambda=10$, $50$, and $100$ nm correspond to
$q=0.0628$, $0.0126$, and $0.00628~\angstrom^{-1}$, respectively.
Near-field optical confinement at the 10-nm scale has been demonstrated
experimentally \cite{Guo2025LSA}.  We use these numbers only to establish an
accessible momentum scale.  The actual $q$ distribution and field enhancement
depend on the launcher geometry, dielectric environment, and optical
frequency.

For the rank-resolved response,
\begin{equation}
\mathcal T_\ell
=
\mathcal C_\ell(\omega)\eta_\ell
\left(\frac{q}{k_0}\right)^{\ell-2}
+
\mathcal O(q^\ell),
\end{equation}
where powers of $k_0$ have been absorbed into the definition of the
coefficient.  The factor plotted in Fig.~\ref{fig:S_experiment}(b),
\begin{equation}
S_\ell=(q/k_0)^{\ell-2},
\end{equation}
must therefore not be interpreted as a prediction of an absolute absorption
fraction.

\subsection{Discrete angular projection}

For one-photon absorption with linear polarization,
$\Delta W(\theta,\phi_q)$ contains polarization harmonics
$m_\theta=0,\pm2$.  Because the polarization axis is unchanged under
$\theta\rightarrow\theta+\pi$, it is sufficient to sample
\begin{equation}
\theta_r=\frac{\pi r}{N_\theta},
\qquad
r=0,\ldots,N_\theta-1.
\end{equation}
The discrete orthogonality relation is
\begin{equation}
\frac{1}{N_\theta}
\sum_{r=0}^{N_\theta-1}
e^{2i\theta_r}
e^{-2is\theta_r}
=
\delta_{1,s\;({\rm mod}\,N_\theta)}.
\end{equation}
The two nonzero polarization harmonics correspond to discrete indices
$\pm1$.  The smallest grid that distinguishes them and the scalar component
is therefore
\begin{equation}
N_\theta^{\rm min}=3.
\label{S:Ntheta}
\end{equation}

After counterpropagating symmetrization,
\begin{equation}
\Delta W^{q{\rm -even}}(\phi_q)
=
\frac{1}{2}
\left[
\Delta W(\phi_q)+\Delta W(\phi_q+\pi)
\right],
\end{equation}
only even momentum harmonics remain and the response is $\pi$-periodic.
Writing
\begin{equation}
m_q=2r_q,
\end{equation}
we sample
\begin{equation}
\phi_s=\frac{\pi s}{N_\phi},
\qquad
s=0,\ldots,N_\phi-1.
\end{equation}
The target rank-resolved harmonic is
\begin{equation}
r_q^\star=\frac{\ell-2}{2}.
\end{equation}
If the response is band-limited to
$|r_q|\leq r_q^\star$, a discrete Fourier transform must contain at least
\begin{equation}
N_\phi^{\rm min}=2r_q^\star+1=\ell-1
\label{S:Nphi}
\end{equation}
points to distinguish $+r_q^\star$ and $-r_q^\star$ without aliasing.
This gives
\begin{equation}
N_\phi^{\rm min}
=
1,\quad3,\quad5
\end{equation}
for $d$-, $g$-, and $i$-wave order.  Equation~(\ref{S:Nphi}) is an ideal band-limited minimum, not a recommended upper bound.  In a real
crystal, higher-order spatial-dispersion harmonics, noise, and imperfect alignment motivate oversampling.

The resulting discrete estimator is
\begin{equation}
\mathcal T_\ell^{\rm disc}
=
\frac{1}{N_\theta N_\phi}
\sum_{r,s}
e^{i[2\theta_r+(\ell-2)\phi_s]}
\Delta W^{q{\rm -even}}
(q;\theta_r,\phi_s).
\label{S:Tdisc}
\end{equation}
In the band-limited limit Eq.~(\ref{S:Tdisc}) is identical to the continuous
angular projection.

\subsection{Finite angular and momentum acceptance}

A real optical mode has a finite distribution of momentum directions.
Let the nominal direction be $\phi_q$ and let the deviation
$\delta\phi$ follow a normalized Gaussian distribution
\begin{equation}
p(\delta\phi)
=
\frac{1}{\sqrt{2\pi}\sigma_\phi}
\exp\!\left[
-\frac{\delta\phi^2}{2\sigma_\phi^2}
\right].
\end{equation}
Convolution of an angular Fourier harmonic gives
\begin{align}
\left\langle
e^{-im_q(\phi_q+\delta\phi)}
\right\rangle
&=
e^{-im_q\phi_q}
\int d(\delta\phi)\,
p(\delta\phi)e^{-im_q\delta\phi}
\nonumber\\
&=
e^{-im_q\phi_q}
e^{-m_q^2\sigma_\phi^2/2}.
\end{align}
For the target $m_q=\ell-2$,
\begin{equation}
\mathcal T_\ell(\sigma_\phi)
=
\mathcal T_\ell(0)
\exp\!\left[
-\frac{(\ell-2)^2\sigma_\phi^2}{2}
\right].
\label{S:angularattenuation}
\end{equation}
A symmetric angular spread therefore reduces the amplitude but introduces no
phase shift and does not change the Fourier index.

A narrow distribution of momentum magnitudes has a similarly benign effect.
For $q=q_c+\delta q$ with
$\langle\delta q\rangle=0$ and variance $\sigma_q^2$,
\begin{equation}
\left\langle q^m\right\rangle
=
q_c^m
\left[
1+
\frac{m(m-1)}{2}
\left(\frac{\sigma_q}{q_c}\right)^2
+
\mathcal O\!\left(
\frac{\sigma_q^4}{q_c^4}
\right)
\right].
\label{S:qspread}
\end{equation}
Thus a sufficiently narrow momentum distribution renormalizes the amplitude
without altering the leading power or angular winding.

\subsection{Experimental background rejection and readout}

A practical acquisition sequence can implement the projections used in the
theory in three logically independent steps.  First, reversing all local
magnetic moments at fixed crystal structure isolates the domain-odd
spin-resolved response.  Second, measurements at counterpropagating
$+\bm q$ and $-\bm q$ remove all odd-$q$ contributions.  Third, the
discrete Fourier transform in Eq.~(\ref{S:Tdisc}) selects the desired
momentum-angle harmonic.

The theory does not require a specific spin detector.  A direct realization
is spin-resolved photoemission, for which imaging spin analyzers and
spin-resolved photoemission microscopy have been demonstrated
\cite{deLaFiguera2017}.  Photoemission electron microscopy has also been used
to characterize strongly confined optical near fields
\cite{Guo2025LSA}.  These observations establish the relevant ingredients
separately; combining a directional high-$q$ optical launcher with
spin-resolved detection for higher-wave altermagnetic tomography remains an
experimental implementation of the present proposal.  Structured-light
photocurrent spectroscopy provides a complementary route
\cite{Hao2026RSI}, while recent theory shows that spatially structured optical
fields naturally generate nonlocal charge and spin-valley responses in
two-dimensional systems \cite{Gunyaga2026PRB}.

\section{First-principles calculations for MnTe}
\label{sec:SM_DFT}

\subsection{Computational details}
\label{sec:SM_DFT_methods}

First-principles calculations were performed with VASP  using the
projector-augmented-wave (PAW) method and the Perdew--Burke--Ernzerhof
generalized-gradient approximation~\cite{Kresse1996VASP,Kresse1999VASP,Blochl1994PAW,PBE1996}.
We used the experimental NiAs structure of $\alpha$-MnTe
(space group $P6_3/mmc$) with $a=4.171$~\angstrom\ and $c=6.686$~\angstrom.  The plane-wave
cutoff was $600$~eV and the self-consistent Brillouin zone was sampled on a
$\Gamma$-centered $12\times12\times8$ mesh~\cite{Monkhorst1976,Takahashi2025npj}.
Electronic iterations were converged to $10^{-6}$~eV or tighter, and Gaussian
smearing of $0.05$~eV was used for the fixed-geometry electronic calculations.

Correlations in the Mn $3d$ shell were treated using the rotationally
invariant DFT+$U$ functional ~\cite{Liechtenstein1995},
with
$U_{\rm Mn}=4.0~\mathrm{eV},\qquad J_{\rm Mn}=0.9~\mathrm{eV}$,
and no Hubbard correction on Te.  The same settings were used in all
collinear, noncollinear, SOC, and dense-$\bm k$ calculations.  The converged
A-type compensated state has antiparallel Mn moments of approximately
$4.4\,\mu_{\rm B}$ and a total moment below the numerical resolution.

 For Relativistic
calculations,  we considered
$\hat{\bm L}_{[11\bar{2}0]}=\left(\frac12,\frac{\sqrt3}{2},0\right)$,
and
$\hat{\bm L}_{[1\bar{1}00]}=\left(\frac{\sqrt3}{2},-\frac12,0\right)$.

\subsection{Non-relativistic momentum-space texture}
\label{sec:SM_DFT_nonSOC}

The nonrelativistic texture was evaluated on an explicit $81\times81$
Cartesian $k_x$--$k_y$ grid at fractional coordinate $k_3=0.45$, giving
$6561$ points.  The two highest valence branches are VASP bands 18 and 19,
with band 19 forming the top valence branch.  We define
\begin{equation}
\Delta E_n(\bm k)=E_{n\uparrow}(\bm k)-E_{n\downarrow}(\bm k).
\label{eq:SM_collinear_split}
\end{equation}
On a circle of fixed $k_\parallel$, the coefficients
\begin{equation}
C_m(k_\parallel,k_3)=\frac{1}{N_\phi}\sum_{j=1}^{N_\phi}
\Delta E(\phi_j)e^{-im\phi_j}
\label{eq:SM_fourier_nonSOC}
\end{equation}
were used to define $A_m=2|C_m|$ for $m>0$ and
\begin{equation}
P_m=\frac{2|C_m|^2}{|C_0|^2+2\sum_{n\ge1}|C_n|^2}.
\label{eq:SM_fourier_power}
\end{equation}
When the selected amplitude vanishes identically, the corresponding purity is
undefined and is left unassigned.

To resolve the out-of-plane dependence, direct-circle calculations were
performed for $k_3=0.40$, $0.45$, $0.475$, $0.490$, $0.500$, $0.510$,
$0.525$, $0.550$, and $0.600$, and for
$k_\parallel=0.10$, $0.18$, and $0.25\,\angstrom^{-1}$.
The signed $C_3$ shown in the Letter is obtained by fixing a common phase
convention on one side of $A$ and projecting all coefficients onto that phase.

\begin{figure*}[t]
\centering
\includegraphics[width=\linewidth]{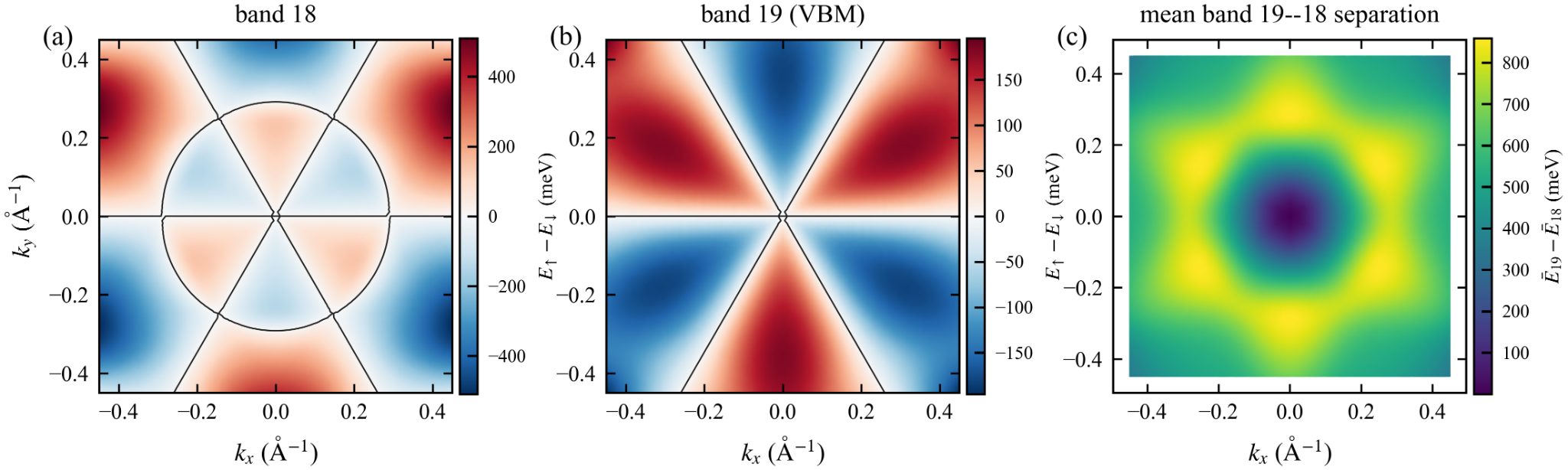}
\caption{\label{fig:S_DFT1}
Band-resolved non-relativistic texture at $k_3=0.45$.
(a),(b) $E_\uparrow-E_\downarrow$ for bands 18 and 19.
Band 18 contains an additional radial node, while the top valence branch retains
a clean six-sector pattern.
(c) Mean separation $\bar E_{19}-\bar E_{18}$, where
$\bar E_n=(E_{n\uparrow}+E_{n\downarrow})/2$. The valence band maximum (VBM) in the non-relativistic calculations is band 19 for both spin-up and spin-down channels.}
\end{figure*}

\begin{figure*}[t]
\centering
\includegraphics[width=\linewidth]{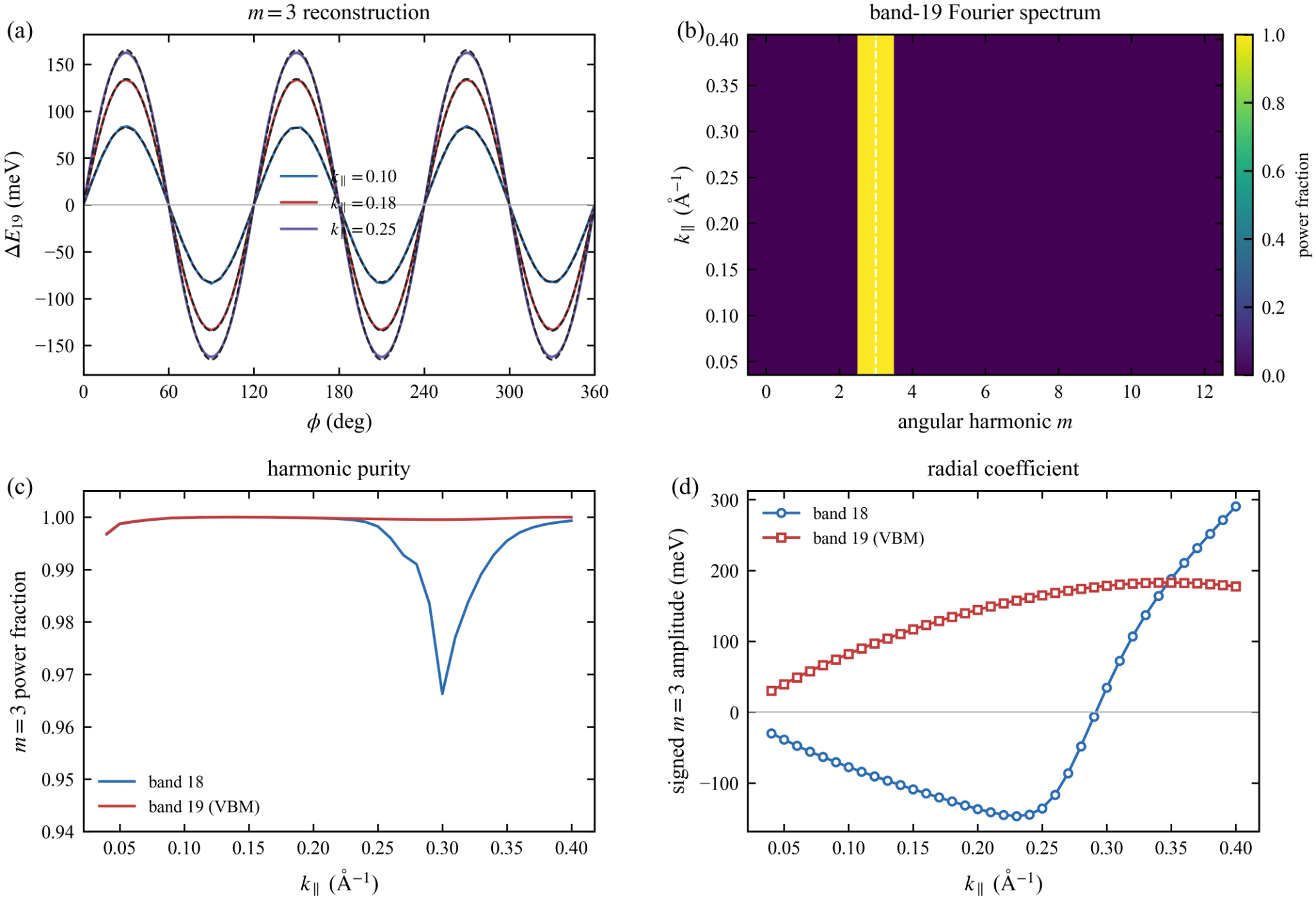}
\caption{\label{fig:S_DFT2}
Fourier characterization at $k_3=0.45$.
(a) Angular traces of the top valence branch and their $m=3$ reconstructions.
(b) Full momentum-angle Fourier spectrum.
(c) $m=3$ purity for bands 18 and 19.
(d) Signed radial $m=3$ coefficients.}
\end{figure*}

\begin{figure*}[t]
\centering
\includegraphics[width=\linewidth]{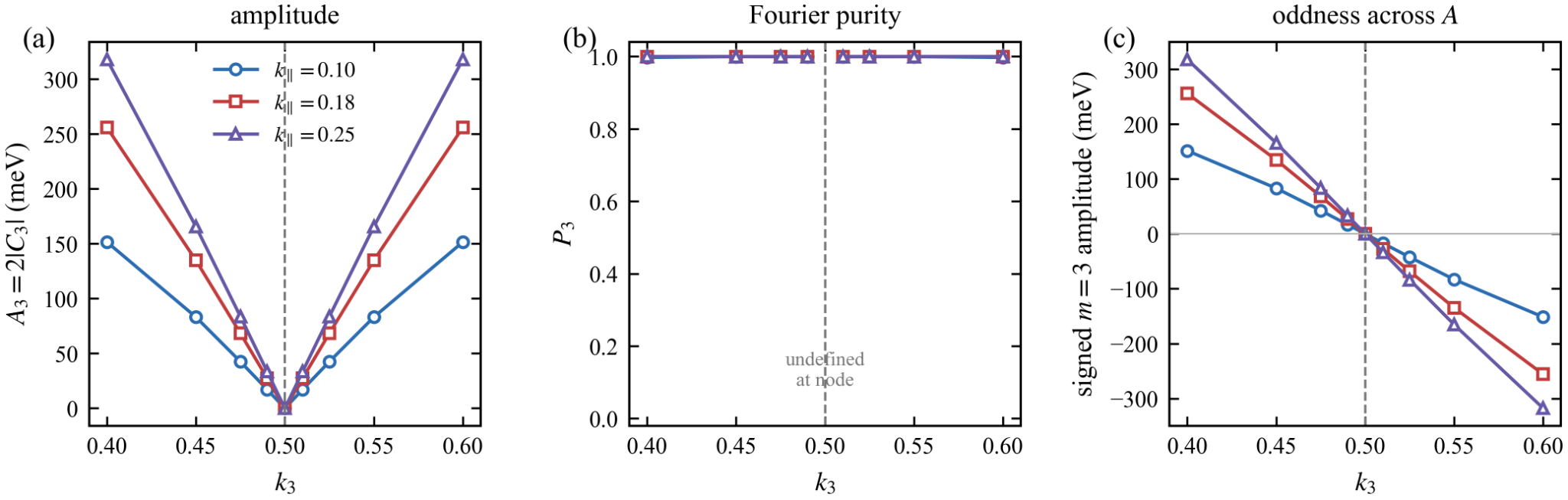}
\caption{\label{fig:S_DFT3}
Evolution through the $A$ plane.
(a) $A_3=2|C_3|$.
(b) Fourier purity; it is undefined at the symmetry-enforced node where
$A_3=0$.
(c) Signed $m=3$ amplitude, which reverses across $k_3=1/2$.}
\end{figure*}

\subsection{SOC topology and selection of the $A$-region planes}
\label{sec:SM_DFT_SOC_topology}

In the relativistic case, band 38 is the top of the valence band, while band 37 has similar dispersion but predominantly opposite Néel-projected spin polarization. 
A 301-point scan along $k_1=k_2=0$ locates the two band-38 maxima at
$k_3=0.44665$ and $0.55341$ for $\bm L\parallel[11\bar{2}0]$, and at
$0.44663$ and $0.55354$ for $\bm L\parallel[1\bar{1}00]$.
Their energies agree within $0.1$--$0.2$~meV, motivating the symmetric
production planes $k_3=0.447$ and $0.553$.

The SOC valence topology at $k_3=0$ was mapped on an $81\times81$ Cartesian
grid.  The highest sampled $\Gamma$-region pockets occur near
$(\pm0.2275,0)\,\angstrom^{-1}$ for $[11\bar{2}0]$ and near
$\pm(0.11375,0.20125)\,\angstrom^{-1}$ for $[1\bar{1}00]$.
They lie approximately $36.7$ and $51.3$~meV, respectively, above the
corresponding axial $A$-region maxima in the present DFT+$U$ description.
Thus SOC strongly reorganizes the band-edge topology near $\Gamma$, while the
$A$-region higher-wave texture is nearly orientation independent.

\begin{figure*}[t]
\centering
\includegraphics[width=\linewidth]{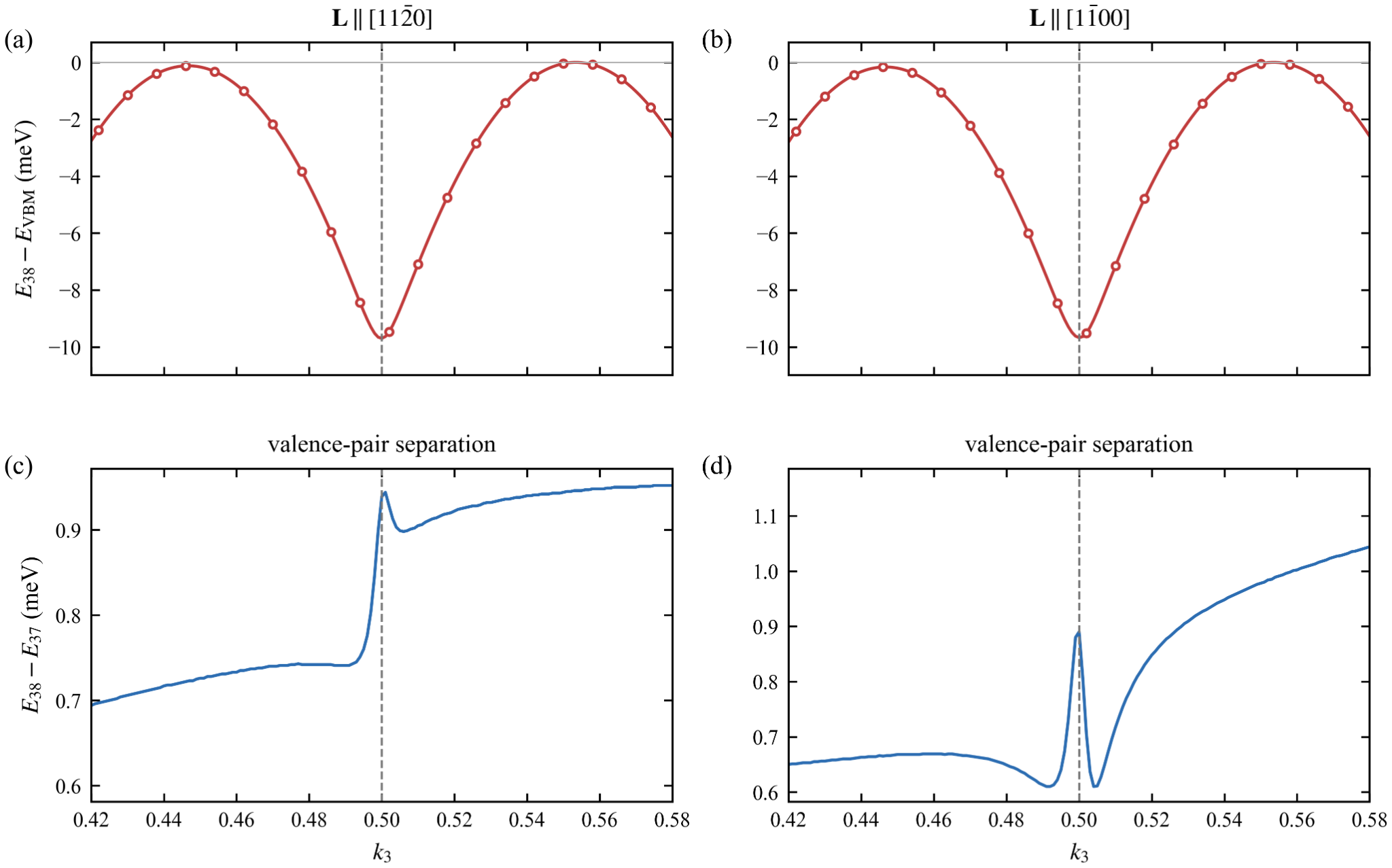}
\caption{\label{fig:S_DFT4}
SOC valence dispersion along $k_1=k_2=0$.
(a),(b) Energy of the upper valence branch relative to its axial maximum E$_{VBM}$ for
the two N\'eel orientations.
(c),(d) Internal separation of the selected valence pair.}
\end{figure*}

\begin{figure*}[t]
\centering
\includegraphics[width=\linewidth]{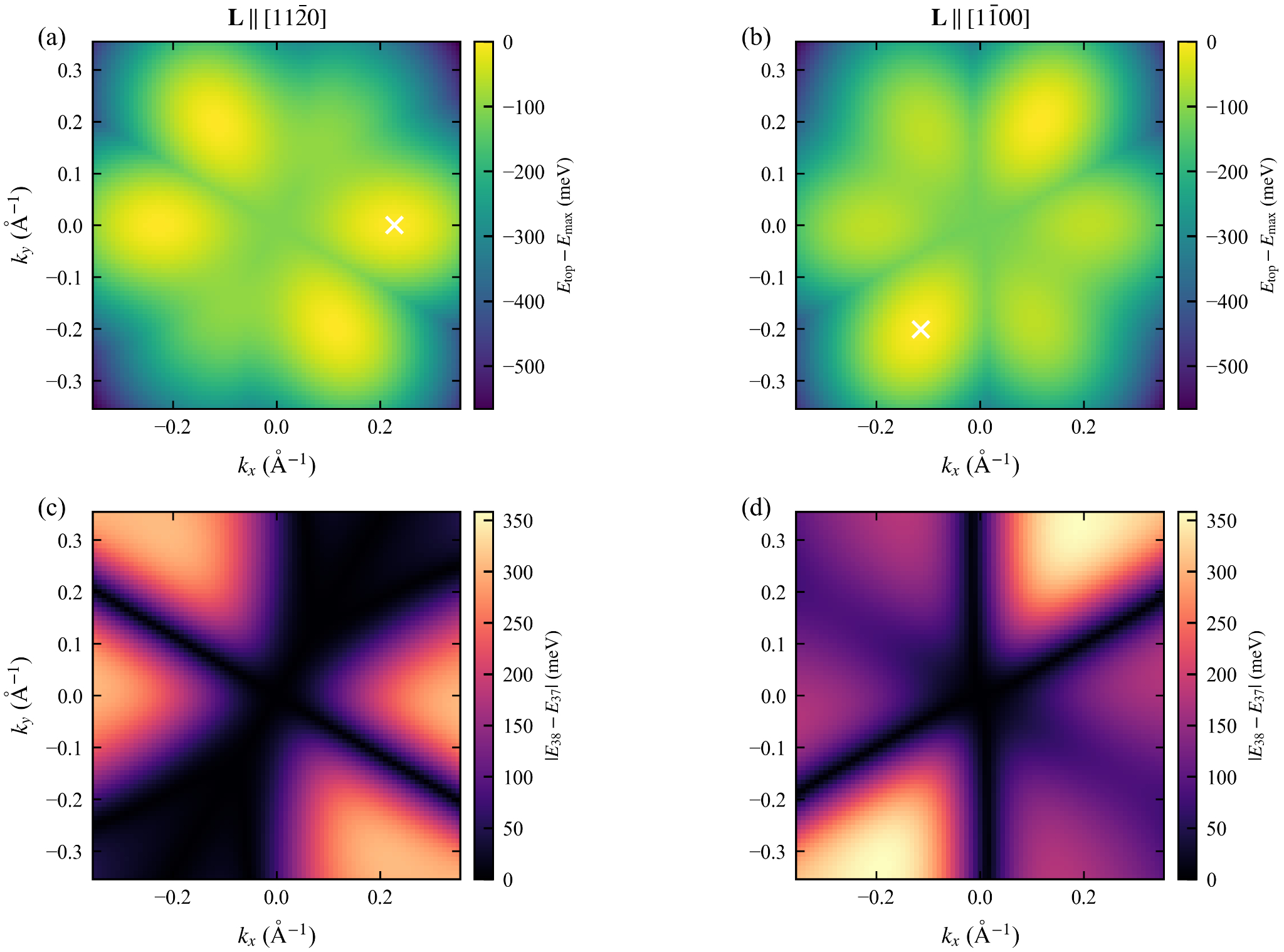}
\caption{\label{fig:S_DFT5}
N\'eel-orientation dependence of the SOC valence topology at $k_3=0$.
(a),(b) Top-valence energy surfaces; crosses mark the numerical maxima.
(c),(d) Corresponding internal valence-pair separation.}
\end{figure*}

\subsection{Relativistic higher-wave texture around $A$}
\label{sec:SM_DFT_SOC_A}

The final SOC sampling used $k_3=0.447,0.553$, radii
$k_\parallel=0.10,0.18,0.25\,\angstrom^{-1}$, and 120 angles per
circle, totaling 720 momenta for each N\'eel orientation.  In the VASP output, the bands 37 and 38 correspond to the lower and upper branches
$v_{\rm l}$ and $v_{\rm u}$ of the selected top-valence pair.
For each branch, the normalized projected spin is
\begin{equation}
\bm s_v^{\rm PAW}=\frac{(m_{vx},m_{vy},m_{vz})}{q_v},
\qquad s_{v,\parallel}=\hat{\bm L}\cdot\bm s_v^{\rm PAW},
\label{eq:SM_spin_projection}
\end{equation}
where $q_v$ is the total PAW projection weight.  We define
\begin{equation}
D_\parallel(\bm k)=\frac{\Delta E_v(\bm k)\,\Delta s_\parallel(\bm k)}{2},
\quad
\Delta E_v=E_{v_{\rm u}}-E_{v_{\rm l}},\quad
\Delta s_\parallel=s_{v_{\rm u},\parallel}-s_{v_{\rm l},\parallel}.
\label{eq:SM_Dparallel}
\end{equation}
This expression is invariant under interchange of the two branch labels and
reduces to the signed spin splitting in the spin-conserving limit.  With SOC, it is a two-band spin-energy correlation based on PAW-projected polarizations.
Its Fourier coefficients are
\begin{equation}
C_m^{\rm SOC}=\frac{1}{N_\phi}\sum_jD_\parallel(\phi_j)e^{-im\phi_j}.
\label{eq:SM_SOC_fourier}
\end{equation}

For $\bm L\parallel[11\bar{2}0]$, the six circles yield
$P_3=0.99939$--$0.99987$; for $\bm L\parallel[1\bar{1}00]$,
$P_3=0.99939$--$0.99986$.  The ratios
$|C_3(0.553)|/|C_3(0.447)|$ lie between $0.99938$ and $1.00006$, and the
phase difference is within $0.041^\circ$ of $180^\circ$.
The mean PAW weights of the two branches are nearly equal and decrease
smoothly from about $0.62$ to $0.59$ over the sampled radii.

\begin{figure*}[t]
\centering
\includegraphics[width=\linewidth]{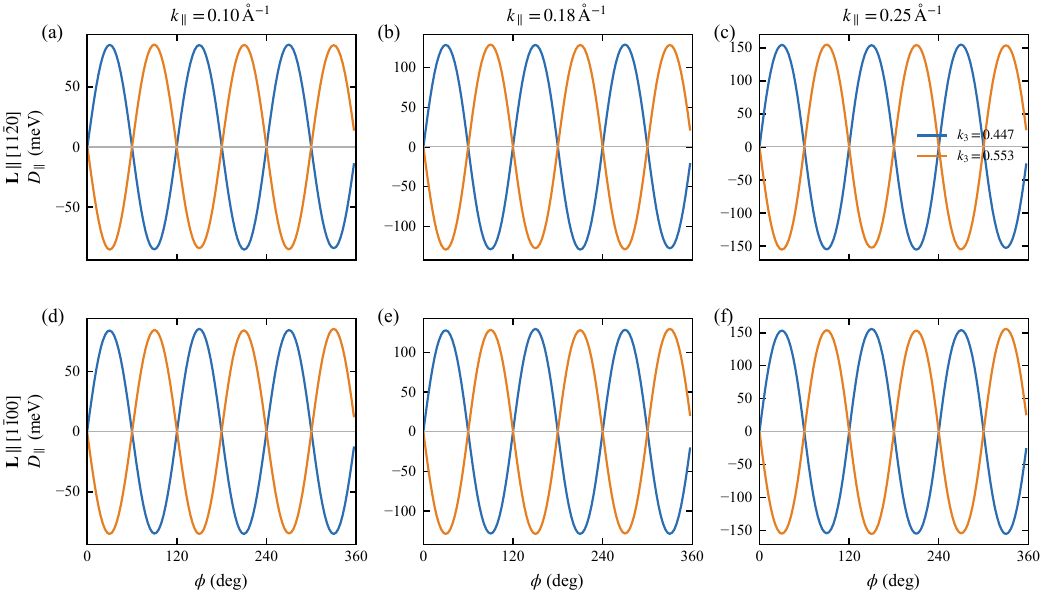}
\caption{\label{fig:S_DFT6}
Complete SOC spin-energy contrast around $A$.
(a)--(c) $\bm L\parallel[11\bar{2}0]$ and
(d)--(f) $\bm L\parallel[1\bar{1}00]$ for the three radii.
Blue and orange denote $k_3=0.447$ and $0.553$; the amplitudes match and the
phases are opposite.}
\end{figure*}

\begin{figure*}[t]
\centering
\includegraphics[width=\linewidth]{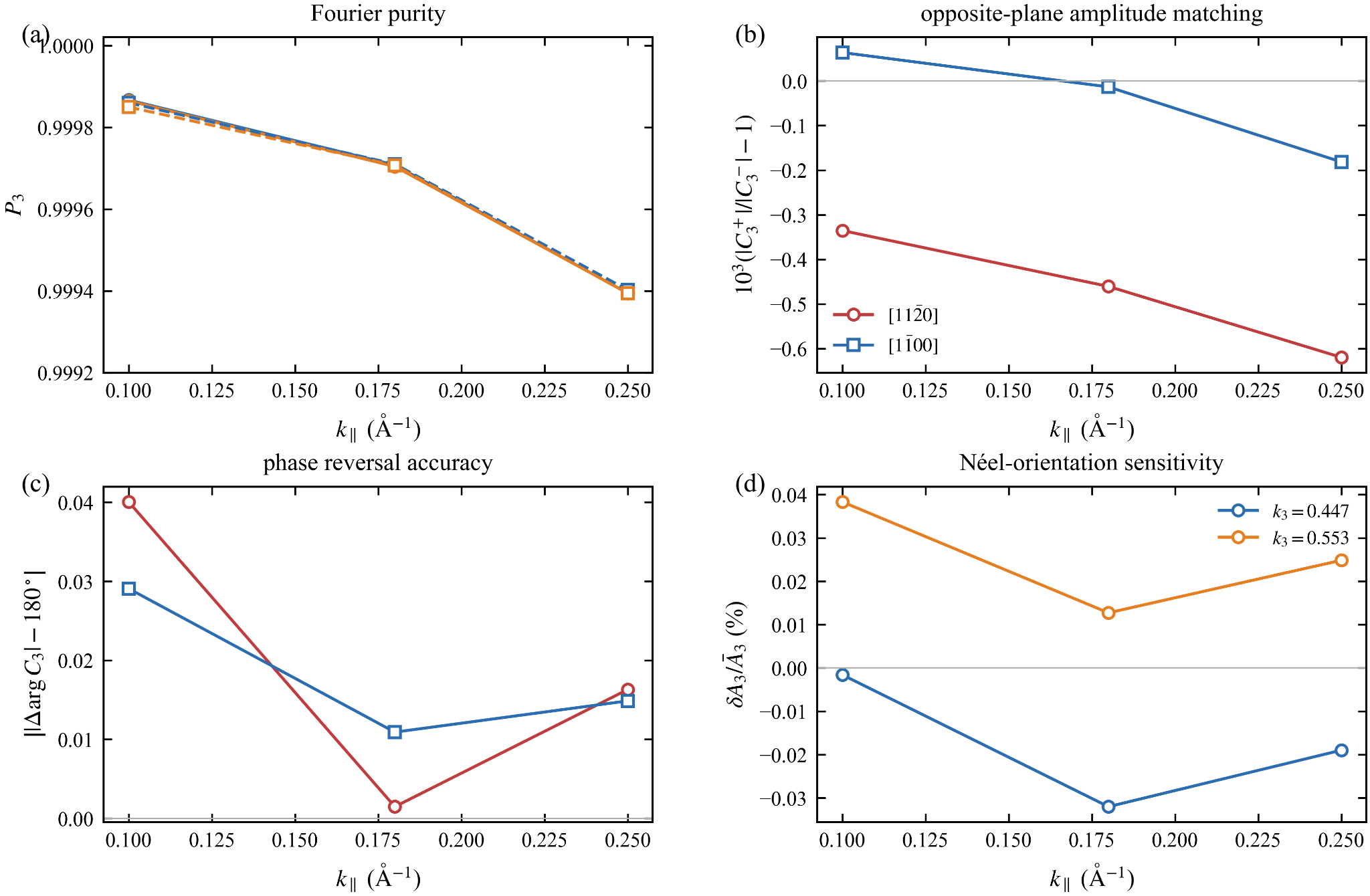}
\caption{\label{fig:S_DFT7}
Quantitative SOC robustness.
(a) $P_3$.
(b) Opposite-plane amplitude mismatch.
(c) Absolute phase-reversal error.
(d) Relative N\'eel-orientation sensitivity
$\delta A_3/\bar A_3$, where
$\delta A_3=A_3^{[1\bar{1}00]}-A_3^{[11\bar{2}0]}$.}
\end{figure*}

\subsection{Isolation of the selected valence manifold}
\label{sec:SM_DFT_isolation}

Using the same 720-point data, we evaluated the minimum separations of the
selected pair from the adjacent VASP bands.  The lower gap
$E_{37}-E_{36}$ remains of order $10^{-1}$~eV or larger, while
$E_{39}-E_{38}$ is of order $1$~eV.  Hence, a nearby crossing with bands 36
or 39 is not responsible for the observed harmonic.

\begin{figure*}[t]
\centering
\includegraphics[width=\linewidth]{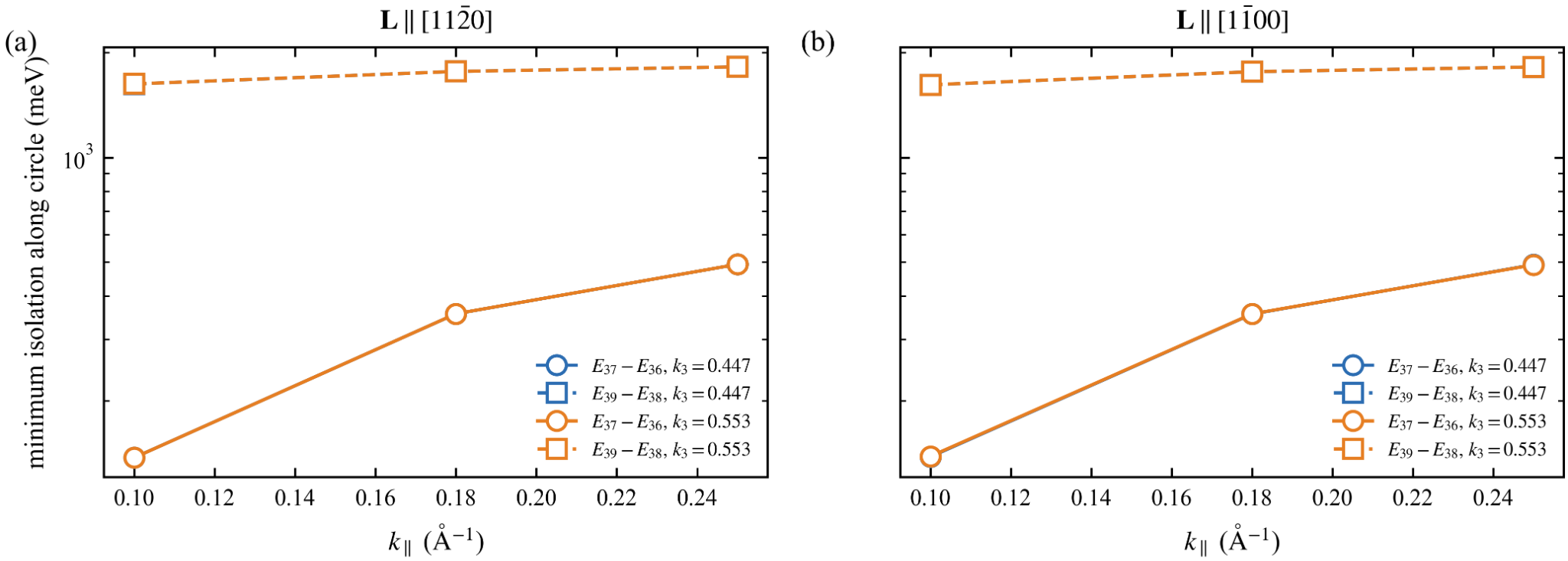}
\caption{\label{fig:S_DFT8}
Isolation of the selected SOC valence pair.
(a),(b) Minimum $E_{37}-E_{36}$ (circles) and
$E_{39}-E_{38}$ (squares) along each direct circle for the two N\'eel
orientations. The two $A$-side planes are distinguished by line style.}
\end{figure*}

\bibliography{references_prl_final}